\documentclass[useAMS,twocolumn,usenatbib]{mn2e}
\usepackage[]{graphicx}

\usepackage{amssymb}
\usepackage{amsfonts}
\usepackage[fleqn]{amsmath}
\usepackage{txfonts}

\usepackage[hyperindex]{hyperref}
\hypersetup{
breaklinks = {true},
colorlinks = {false},
linkcolor={black},
pdfpagemode = {None}, 
pdfborder = {0 0 1},
pdftitle = {Measuring primordial non-Gaussianity with weak-lensing surveys},
pdfsubject = {Expected constraints on primordial non-Gaussianity of the local-type from weak-lensing surveys},
pdfauthor = {Stefan Hilbert, Laura Marian, Robert. E. Smith, and Vincent Desjacques},
pdfkeywords = {gravitational lensing: weak -- large-scale structure of the Universe -- cosmological parameters -- early Universe -- inflation --methods: numerical}
}

\allowdisplaybreaks[4]
\bibpunct{(}{)}{;}{a}{}{,}

\def\figrelpath{}

\def\apj{ApJ}%
\def\apjl{ApJ}%
\def\apjs{ApJS}%
\def\aap{A\&A}%
\def\mnras{MNRAS}%
\def\prd{Phys.~Rev.~D}%
\def\physrep{Phys.~Rep.}%
\def\jcap{JCAP}%

\newcommand{\satellitename}[1]{\textit{#1}}
\newcommand{\telescopename}[1]{{#1}}
\newcommand{\softwarename}[1]{\textsc{#1}}

\newcommand{\vect}[1]{\boldsymbol{#1}}
\newcommand{\matr}[1]{\mathsf{#1}}
\newcommand{\matrb}[1]{\boldsymbol{\mathsf{#1}}}

\newcommand{\transposed}[1]{#1^\mathsf{t}}

\newcommand{\ii}{\mathrm{i}}
\newcommand{\ee}{\mathrm{e}}
\newcommand{\dd}{\mathrm{d}}

\newcommand{\R}{\mathbb{R}}

\newcommand{\parder}[3][]{\frac{\partial^{#1} {#2}}{\partial {#3}^{#1}}}
\newcommand{\dparder}[3][]{\dfrac{\partial^{#1} {#2}}{\partial {#3}^{#1}}}

\newcommand{\totder}[3][]{\frac{\mathrm{d}^{#1} {#2}}{\mathrm{d} {#3}^{#1}}}

\newcommand{\ttotder}[3][]{{\mathrm{d}^{#1} {#2}/\mathrm{d} {#3}^{#1}}}
\newcommand{\diff}[2][]{\mathrm{d}^{#1}{#2}}
\newcommand{\idiff}[2][]{\!\!\mathrm{d}^{#1}{#2}}

\newcommand{\EV}[1]{\left\langle{#1}\right\rangle}
\newcommand{\bEV}[1]{\bigl\langle{#1}\bigr\rangle}

\newcommand{\est}[1]{\hat{#1}}

\newcommand{\ft}[1]{\tilde{#1}}

\newcommand{\kpc}{\ensuremath{\mathrm{kpc}}}
\newcommand{\Mpc}{\ensuremath{\mathrm{Mpc}}}

\newcommand{\Msolar}{\ensuremath{\mathrm{M}_\odot}}

\newcommand{\arcsect}{\ensuremath{\mathrm{arcsec}}}
\newcommand{\arcmint}{\ensuremath{\mathrm{arcmin}}}
\newcommand{\degt}{\ensuremath{\mathrm{deg}}}

\newcommand{\vtheta}{\vect{\theta}}
\newcommand{\vvartheta}{\vect{\vartheta}}

\newcommand{\gammat}{\gamma_{\mathrm{t}}}
\newcommand{\gammax}{\gamma_{\times}}

\newcommand{\zmedian}{z_\text{median}}

\newcommand{\ngal}{n_{\text{g}}}

\newcommand{\sigmaepsilongalpc}{\sigma_{\epsilon_i}}

\newcommand{\sigmapixel}{\sigma_{\text{pix}}}
\newcommand{\Apixel}{A_{\text{pix}}}

\newcommand{\fNL}{f_\text{NL}}
\newcommand{\Map}{M_\text{ap}}
\newcommand{\dpa}{\phi_{M}}
\newcommand{\Np}{N_{M}}

\newcommand{\Model}{\mathcal{M}}
\newcommand{\Nparam}{N_{\text{p}}}
\newcommand{\Ndata}{N_{\text{d}}}
\newcommand{\Nreal}{N_{\text{r}}}

\newcommand{\AllowedRegion}{S}

\newcommand{\prior}{p}
\newcommand{\posterior}{p}
\newcommand{\likelihood}{L}
\newcommand{\loglikelihood}{\ln L}

\DeclareMathOperator{\cov}{Cov}

\newcommand\Eqn[1]{Eq.\,(\ref{#1})}

\newcommand{\nn}{\nonumber}

\newcommand{\dk}{{\dd^3\vect{k}}}

\newcommand{\TT}{{\mathrm{TT}}}
\newcommand{\EE}{{\mathrm{EE}}}
\newcommand{\TE}{{\mathrm{TE}}}

\begin{document}

\title[Measuring primordial non-Gaussianity with weak-lensing surveys]{Measuring primordial non-Gaussianity with weak-lensing surveys}
\author[S. Hilbert, L. Marian, R. E. Smith, and V. Desjacques]{
Stefan Hilbert$^{1,2}$\thanks{\texttt{hilbert@stanford.edu}},
Laura Marian$^{2}$,
Robert E. Smith$^{2,3}$, and
Vincent Desjacques$^{3,4}$
\\$^{1}$Kavli Institute of Particle Astrophysics and Cosmology (KIPAC), Stanford University, 452 Lomita Mall, Stanford, CA 94305, and \\
SLAC National Accelerator Laboratory, 2575 Sand Hill Road, M/S 29, Menlo Park, CA 94025
\\$^{2}$Argelander-Institut f{\"u}r Astronomie, Auf dem H{\"u}gel 71, 53121 Bonn, Germany
\\$^{3}$Institute for Theoretical Physics, University of Z{\"u}rich, Z{\"u}rich, CH 8057, Switzerland
\\$^{4}$Universite de Geneve and Center for Astroparticle Physics, 24 Quai Ernest Ansermet, 1211 Geneve 4, Switzerland
}
\date{\today}
\maketitle

\begin{abstract}
Measuring the non-Gaussianity of the initial matter density fluctuations may provide powerful insights into cosmic inflation and the origin of structures. Although current information on primordial non-Gaussianity comes mostly from the Cosmic Microwave Background, obtaining constraints from lower-redshift observables will be an important task for future surveys. We study the ability of future weak lensing (WL) surveys to constrain primordial non-Gaussianity of the local type. We use a large ensemble of simulated WL maps with survey specifications relevant to \satellitename{Euclid} and LSST. The simulations assume Cold Dark Matter cosmologies that vary certain parameters around fiducial values: the non-Gaussianity parameter $\fNL$, the matter density parameter $\Omega_{\text{m}}$, the amplitude of the matter power spectrum $\sigma_8$, the spectral index of the primordial power spectrum $n_{\text{s}}$, and the dark-energy equation-of-state parameter $w_0$. We assess the sensitivity of the cosmic shear correlation functions, the third-order aperture mass statistics, and the abundance of shear peaks to these parameters. We find that each of the considered probes provides unmarginalized constraints of $\Delta\fNL \sim 20$ on $\fNL$. Marginalized constraints from any individual WL probe are much weaker due to strong correlations between parameters. However, the parameter errors can be substantially reduced by combining information from different WL probes. Combining all WL probes yields the following marginal (68\% confidence level) uncertainties: $\Delta \fNL \sim 50$, $\Delta \Omega_\text{m} \sim 0.002$, $\Delta \sigma_8 \sim 0.004$, $\Delta n_\text{s} \sim 0.007$, and $\Delta w_0 \sim 0.03$. We examine the bias induced by neglecting $\fNL$ on the constraints on the other parameters. We find $\sigma_8$ and $w_0$ to be the most affected. Moreover, neglecting non-Gaussianity leads to a severe underestimation of the  uncertainties in the other cosmological parameters. We conclude that a full exploitation of future WL surveys requires a joint analysis of different WL probes. Furthermore, if not taken into account, a non-vanishing level of primordial non-Gaussianity will bias the estimated cosmological parameters and uncertainties for future surveys.
\end{abstract}


\section{Introduction}

According to Big Bang cosmologies, the rich matter structures seen today in our Universe have grown from tiny density fluctuations at very early times. There are a number of competing theories for the physical origins of these primordial density perturbations, e.g. the Ekpyrotic model \citep{Khouryetal2001} or the string gas cosmology \citep{Brandenbergeretal2007}. The leading explanation is the inflationary paradigm \citep{MukhanovChibisov1981}. The standard single-field slowly-rolling inflationary model provides four fundamental predictions: a nearly flat spatial curvature for spacetime, a nearly scale-invariant power spectrum of primordial curvature perturbations, a nearly Gaussian distribution of primordial density perturbations, and a scale-invariant spectrum of gravitational waves. The first three predictions have been tested by observations of the cosmic microwave background radiation (CMB) with the \satellitename{Wilkinson Microwave Anisotropy Probe} \citep[\satellitename{WMAP},][]{KomatsuEtal2011short}. The discovery of primordial gravitational wave signatures in the CMB is one of the main goals for the \satellitename{Planck} mission \citep{PlanckBlueBook}.

Recently much attention has been turned towards using observations of the statistical properties of the large-scale structure to further falsify models for the origin of density perturbations. For example, single field inflation leads to a near-Gaussian distribution of fluctuations \citep{AcquavivaEtal2003,Maldacena2003}. Hence, if significant departures from Gaussianity are detected, all single-field inflationary models could be ruled out \citep{CreminelliZaldarriaga2004}. In such models, departures from Gaussianity can be characterized by the local expansion of the primordial potential perturbations \citep{MatarreseVerdeJimenez2000}:
\begin{equation}
\label{eq:phi_local_non_gaussianity}
\Phi(\vect{x}) = \phi(\vect{x}) + \fNL  \left[ \phi^2(\vect{x}) - \EV{\phi^2(\vect{x})}\right]
,
\end{equation}
Here, $\phi(\vect{x})$ denotes the Gaussian potential perturbation at spacetime point $\vect{x}$ after matter radiation equality, and the parameter $\fNL$ quantifies the deviation from Gaussianity. In linear theory, typical fluctuations are very small with $\phi \sim 10^{-5}$. Thus, for $|\fNL|<10^5$, the typical non-Gaussian corrections, are small, too: $(\Phi-\phi)/\phi\sim \fNL \phi$.

Currently, the best constraints on local-type primordial non-Gaussianity come from the CMB bispectrum in the seven-year \satellitename{WMAP} data \citep{KomatsuEtal2011short}: \mbox{$\fNL = 32 \pm 21$} at 68\% confidence level (CL). Though not widely anticipated a few years ago, it has recently been shown that probes of structures at lower redshifts could also provide competitive constraints, in particular the nonlinear matter power spectrum \citep{Seto2001,AmaraRefregier2004,DesjacquesSeljakIliev2009,PillepichPorcianiHahn2010,SmithDesjacquesMarian2011}, the abundance of massive clusters \citep{MatarreseVerdeJimenez2000,LoVerdeEtal2008}, the galaxy bispectrum \citep{SefusattiKomatsu2007,NishimichiEtal2010,BaldaufSeljakSenatore2011,SefusattiCrocceDesjacques2012}, and the scale-dependence of galaxy clustering \citep{Dalaletal2008,DesjacquesSeljakIliev2009,PillepichPorcianiHahn2010}.

In this paper, we focus on the question: How well can a large weak lensing (WL) survey constrain the level of primordial non-Gaussianity in the initial conditions? This is of prime concern given upcoming surveys like the Kilo Degree Survey \citep[KiDS, $\sim1500\,\degt^2$,][]{Kuijken2007} and the Dark Energy Survey \citep[DES, $\sim5500\,\degt^2$,][]{DES2005}, or planned full sky surveys such as those with \satellitename{Euclid} \citep[$\sim15\,000\,\degt^2$,][]{LaureijsEtal2011_Euclid_DSR} or the \telescopename{Large Synoptic Survey Telescope} \citep[\telescopename{LSST},$\sim20\,000\,\degt^2$,][]{LSST_Science_Book_2009}. 

There have been various forecasts for the expected error $\Delta\fNL$ on the non-Gaussianity parameter $\fNL$ from future weak lensing surveys. For example, \citet{JeongKomatsuJain2009} predicted that $\Delta\fNL \approx 50$ could be obtained from galaxy-galaxy lensing on large angular scales. \citet{FedeliMoscardini2010} found that $\Delta\fNL<$ a few tens may be possible from a $\chi^2$-analysis of the convergence power spectrum. Using a halo model, \citet{MaturiFedeliMoscardini2011} obtained $\Delta\fNL\sim30$ for a combined analysis of the convergence power spectra and the abundance of shear peaks in a \satellitename{Euclid}-like survey. \citet{SchaeferEtal2012} predicted that $\Delta\fNL \approx 200$ could be achieved from analyzing convergence bispectra, which could be further reduced to $\Delta\fNL \approx 30$ if sufficiently tight priors on all other relevant cosmological parameters are available. \citet{GiannantonioEtal2012} found that a combined analysis of weak lensing and galaxy clustering could give $\Delta\fNL < 10$.
\citet{OguriTakada2011} estimated $\Delta \fNL \sim 10$ for an imaging survey combining cluster counts, the cluster correlation function, and stacked WL signals of clusters.

Whilst this is encouraging, all these studies rely heavily upon approximations for the nonlinear matter power spectrum, bispectrum and analytic modelling of shear peak abundances that are not well tested or understood in non-Gaussian models. Furthermore, the gain in information from combining different probes has not yet been fully investigated. Instead, approximations of various degrees of plausibility have been made for the cross-covariances between different observables. Moreover, most of these forecasts ignore degeneracies of $\fNL$ with other cosmological parameters. Thus, one should be circumspect when interpreting these forecasts.

In this paper we take a more direct approach to answering this question. We use an ensemble of $N$-body simulations of structure formation in flat Dark-Energy Cold-Dark-Matter ($w$CDM) cosmologies, through which we perform full gravitational ray-tracing to construct a suite of large mock weak lensing maps (e.g. the maps for our fiducial cosmological model cover a total area of $14\,000\,\degt^2$). Initial results from these simulations were already presented in \citet{MarianEtal2011}, where we showed that the abundance of shear peaks was sensitive to $\fNL$, and that one may obtain an error of $\Delta\fNL \approx 13$ when all other cosmological parameters are known.
In this work we present a much wider study about combining information from the cosmic-shear two-point correlations $\xi_+$ and $\xi_-$, the aperture mass statistics $\bEV{\Map^3}$, and the abundance of shear peaks $\dpa$. Our approach allows us to fully explore the cross-correlations between these different weak lensing probes.
Furthermore, the initial suite of simulations used in \citet{MarianEtal2011} are augmented with cosmological models that take into account variations in other cosmological parameters.
Thus, we can investigate the joint constraints from WL in the parameter space spanned by: the non-Gaussianity parameter $\fNL$, the mean matter density $\Omega_\text{m}$, normalization $\sigma_8$ and spectral index $n_\text{s}$ of the matter power spectrum, and the equation-of-state (EOS) parameter $w_0$ of the dark energy (DE).

Ray-tracing simulations have been used recently by \citet{PaceEtal2011} and \citet{FedeliEtal2011} to study the effects of primordial non-Gaussianity on basic lensing quantities and the skeleton of the convergence field. In contrast to their work, we focus in our work on observables more easily accessible in weak lensing surveys. Moreover, we include realistic levels of noise directly in our simulations for the error analysis.
Finally, our larger set of simulations allows us to explore a higher-dimensional parameter space in a uniform manner. 

The paper is broken down as follows:
In Section \ref{sec:theory}, we briefly discuss the weak lensing cosmological probes that we employ in this study.
In Section \ref{sec:methods}, we give an overview of the cosmological $N$-body simulations and the ray-tracing, and review how we estimate the observables and their covariance from the simulated data.
In Section \ref{sec:results}, we consider the effects of primordial non-Gaussianity on the weak lensing observables. 
We also discuss the expected statistical uncertainties and covariances of the weak lensing observables as well as their dependence on the full set of considered cosmological parameters.
Furthermore, we present forecasts for unmarginalized and marginalized constraints on cosmological parameters, and discuss possible biases arising from neglecting finite non-Gaussianity.
The paper concludes with a summary and discussion in Section \ref{sec:summary}.

\section{Theory}
\label{sec:theory}

\subsection{Gravitational lensing}
\label{sec:theory:gravitational_lensing}

Gravitational lensing is the process whereby photon bundles emitted from distant light sources are deflected by the gravitational field of matter structures along the line of sight. These deflections can thus magnify, shift and shear the images of distant galaxies \citep[e.g.][]{BartelmannSchneider2001_WL_review,SchneiderKochanekWambsganss_book}. The observed image position $\vect{\theta}=(\theta_1,\theta_2)$ of a source at redshift $z$ may thus differ from the source's `true' angular position $\vect{\beta}=\bigl(\beta_1(\vect{\theta},z),\beta_2(\vect{\theta},z)\bigr)$. Image distortions caused by differential deflection can be quantified locally by the distortion matrix
\begin{equation}
\label{eq:lens_distortion}
 \left(\parder{\beta_i(\vect{\theta},z)}{\theta_j}\right)_{i,j=1,2} =
\begin{pmatrix}
 1 - \kappa - \gamma_1 & - \gamma_2 - \omega \\
 - \gamma_2 + \omega &  1 - \kappa + \gamma_1
\end{pmatrix},
\end{equation}
whose decomposition defines the convergence $\kappa(\vect{\theta},z)$, the complex shear $\gamma(\vect{\theta},z)= \gamma_1(\vect{\theta},z) + \ii \gamma_2(\vect{\theta},z)$, and the asymmetry $\omega(\vect{\theta},z)$. In weak lensing, the asymmetry $\omega$ is usually small enough to be neglected \citep[e.g.][]{HilbertEtal2009_RT,KrauseHirata2010}. In that case, the convergence $\kappa$ and shear $\gamma$ are related by a simple phase factor in Fourier space. Thus, in weak lensing, the shear field contains the same information as the convergence.

To a good approximation, the convergence is given by a projection of the matter density along the line of sight weighted by a geometrical factor, which depends on the distances between observer, deflecting matter structure, and source. The convergence field thus bears information on the growth of structures as well as on the large-scale geometry of the Universe.

For a population of sources with known redshift distribution $p_z(z)$, an effective convergence and shear can be defined by
\begin{align}
\label{eq:df_effective_convergence}
  \kappa(\vect{\theta}) &= \int\idiff[]{z}\, p_z(z) \kappa(\vect{\theta}, z) 
  \quad\text{and}\\
\label{eq:df_effective_shear}
  \gamma(\vect{\theta}) &= \int\idiff[]{z}\, p_z(z) \gamma(\vect{\theta}, z).
\end{align}
Furthermore, the shear $\gamma(\vtheta)$ at a position $\vtheta$ can be decomposed into a tangential component $\gammat(\vtheta,\vvartheta)$ and a cross component $\gammax(\vtheta,\vvartheta)$ with respect to any given direction $\vvartheta$ \citep[e.g.][]{SchneiderVanWaerbekeMellier2002}:
\begin{align}
\label{eq:df_effective_shear_tangential_component}
\gammat(\vtheta,\vvartheta) &= -\Re\!\left[\!\gamma(\vtheta)\ee^{-2\ii\varphi(\vvartheta)}\!\right]
  \quad\text{and}\\
\label{eq:df_effective_shear_cross_component}
\gammax(\vtheta,\vvartheta) &= -\Im\!\left[\!\gamma(\vtheta)\ee^{-2\ii\varphi(\vvartheta)}\!\right]
,
\end{align}
where $\varphi(\vvartheta)$ denotes the polar angle of the vector $\vvartheta$.

\subsection{Weak lensing observables}
\label{sec:theory:weak_lensing_observables}

Owing to the fact that we do not know the intrinsic shapes of individual background galaxies, the quantities $\kappa$ and $\gamma$ are not direct observables. Nevertheless, we can use correlations in the shapes of the background galaxies to construct quantities that are observable.

\subsubsection{Cosmic shear correlation functions}
\label{sec:theory:weak_lensing_observables:cosmic_shear}

In this study, the first set of lensing observables that we consider are the cosmic shear two-point correlation functions \citep[e.g.][]{Kaiser1992}:
\begin{align}
  \xi_{\pm} \bigl(\lvert\vvartheta \rvert\bigr) &=
  \EV{\gammat(\vtheta,\vvartheta)\gammat(\vtheta+\vvartheta,\vvartheta)} \pm \EV{\gammax(\vtheta,\vvartheta)\gammax(\vtheta+\vvartheta,\vvartheta)}
,
\end{align}
where the angle brackets $\bEV{\dots}$ denote the expectation for an ensemble of universes constrained to have a particular set of cosmological parameters. The cosmic shear correlation functions $\xi_{+}$ and $\xi_{-}$ can be directly estimated from the observed ellipticities of the images of distant galaxies.

In weak lensing, the shear correlation functions can be written in terms of the convergence power spectrum \citep{Kaiser1992}:
\begin{equation}
\xi_{+/-}(\vartheta)
=
\int_0^\infty
\frac{\dd \ell \ell}{2\pi}
J_{0/4}(\ell \vartheta)
P_{\kappa}(\ell)
,
\end{equation}
where $J_n(x)$ denotes the Bessel function of the first kind, and $P_{\kappa}(\ell)$ denotes the angular power spectrum of convergence fluctuations.

\subsubsection{Aperture mass statistics}
\label{sec:theory:weak_lensing_observables:aperture_mass}

The aperture mass $\Map(\vtheta, \vartheta)$ at position $\vtheta$ and scale $\vartheta$ can be defined as a filtered version of convergence or shear field \citep[][]{SchneiderEtal1998_Map}:
\begin{equation}
\begin{split}
  \Map(\vtheta, \vartheta) 
  &=
    \int\diff[2]{\vtheta'} U(\lvert \vtheta' - \vtheta \rvert, \vartheta) \kappa(\vtheta')
  \\&=
   \int\diff[2]{\vtheta'} Q(\lvert \vtheta' - \vtheta \rvert, \vartheta) \gammat(\vtheta', \vtheta' - \vtheta),
 \\&=
  - \Re \left[ \int\diff[2]{\vtheta'} Q(\lvert \vtheta' - \vtheta \rvert, \vartheta) \ee^{-2 \ii \varphi(\vtheta' - \vtheta)} \gamma(\vtheta') \right]
\end{split}
\end{equation} 
where
\begin{equation}
  Q(\theta, \vartheta) = \frac{2}{\theta^2}\int_0^{\theta}\diff[]{\theta'} \theta'  U(\theta', \vartheta) - U(\theta, \vartheta)
.
\end{equation}
Here, we use the third-order aperture mass statistics $\bEV{\Map^3}(\vartheta) = \bEV{\Map(\vtheta,\vartheta)^3}$ with the filter functions
\begin{subequations}
\label{eq:M_ap_filter_functions}
\begin{align}
\label{eq:M_ap_filter_function_U}
  U(\theta, \vartheta) &= \frac{1}{2\pi \vartheta^2}\left( 1 - \frac{\theta^2}{2 \vartheta^2} \right) \exp\left(\frac{-\theta^2}{2\vartheta^2} \right)
  \quad\Leftrightarrow\\
\label{eq:M_ap_filter_function_Q}
  Q(\theta, \vartheta) &= \frac{\theta^2}{4\pi \vartheta^4} \exp\left(\frac{-\theta^2}{2\vartheta^2} \right)
\end{align}
\end{subequations}
as another lensing probe of cosmology. The statistics $\bEV{\Map^3}(\vartheta)$ can also be estimated in a straightforward manner from the observed image ellipticities of distant galaxies \citep[][]{SchneiderEtal1998_Map}.

In the limit of weak lensing, the expectation value of $\Map^3$ is related to the convergence bispectrum $B_{\kappa}$ by \citep[e.g.][]{SchneiderKilbingerLombardi2005}:
\begin{equation}
\begin{split}
\bEV{\Map^3}(\vartheta)
&=
24\pi
\!\!\int_0^\infty\!\! \dd \ell \ell
\!\!\int_{\ell/2}^{\ell}\!\! \dd \ell' \ell'
\!\!\int_0^{\arccos\left(\frac{\ell}{2\ell'}\right)}\!\!\dd \varphi
\\&\quad\times
\ft{U}(\ell,\vartheta) \ft{U}(\ell',\vartheta) \ft{U}(\ell'',\vartheta)
B_{\kappa}(\ell,\ell',\ell'')
,
\end{split}
\end{equation}
where $\ell'' = \sqrt{\ell^2 + \ell'^2 - 2 \ell \ell' \cos\varphi}$, and $\ft{U}(\ell, \vartheta)$ denotes the (two-dimensional) Fourier transform of the filter $U(\theta, \vartheta)$.

\subsubsection{Shear peak abundance}
\label{sec:theory:weak_lensing_observables:shear_peak_abundance}

The observed shear field can be used to search for massive galaxy groups and clusters by applying a filter function matched to their expected lensing signal \citep[e.g.][]{HamanaTakadaYoshida2004}. Here, we consider the hierarchical filtering method \citep[][]{MarianSmithBernstein2009,MarianEtal2011,MarianEtal2012} with aperture-mass filters following the expected lensing signal of the Navarro-Frenk-White (NFW) profiles \citep[][]{NavarroFrenkWhite1997}. The observed shear field is convolved with a series of filters $Q_\text{m}(\vtheta;M)$ of different size, from the largest down to the smallest. The size is parametrized by the filter mass $M$:
\begin{equation}
  Q_\text{m}(\vtheta;M) = \mathcal{C}(M) \gamma_{\text{m}}(\vtheta;M),
\end{equation}
where $\gamma_{\text{m}}(\vtheta;M)$ is the shear signal expected for an NFW halo of virial mass $M$ at a model redshift $z_\text{m} = 0.3$ (which is roughly where halos are most effective lenses for the considered source redshifts). These weights (approximately) maximize the signal-to-noise ratio for the corresponding peaks \citep{MarianSmithBernstein2009,MarianEtal2011}. The normalization $\mathcal{C}(M)$ is chosen such that the expected filtered signal of a halo of mass $M$ at the model redshift $z_\text{m}$ equals the halo mass $M$.

All maxima in the filtered shear maps are considered as potential shear peaks. Several selection criteria are used to decide whether to regard a maximum indeed as shear peak: i) The maximum must have a signal-to-noise ratio larger than a particular threshold. ii) Its amplitude $\Map$ must be equal or larger than the mass of the filter $M$ used in the smoothing. iii) Its position must have a certain minimum separation to other peaks with larger amplitude.

A peak will appear with different amplitudes in maps smoothed with filters of different size. To assign a unique peak amplitude to a peak at position $\vtheta$, we seek the filter mass $M$ that matches the peak amplitude $\Map(\vtheta;M)$, i.e. we solve the equation $\Map (\vtheta;M) = M$ for $M$ \citep[see][for details]{MarianEtal2012}. This removes the arbitrariness in the choice of filter scale.
Moreover, the resulting filter size also maximizes the signal-to-noise ratio $S/N$ for an NFW halo of virial mass $M$ at the model redshift $z_\text{m} = 0.3$ thanks to our particular definition of the filter amplitude. This should also (roughly) maximize the sensitivity of the shear peak counts to peaks caused by real structures, and thereby to cosmological parameters. 
In the rest of this text, when we refer to the `mass' of a peak, we shall mean the filter mass $M$ for which equation $\Map=M$ is satisfied.

The number of observed peaks in the filtered fields contains information on the abundance of massive structures, and thus on the cosmology \citep[e.g.][]{MarianBernstein2006,DietrichHartlap2010}. We define the differential peak abundance $\dpa(M)$ at filter mass $M$ is by:
\begin{equation}
\begin{split}
\dpa(M) = - \frac{\partial \Np(M, A)}{A \; \partial \log_{10}(M)}
,
\end{split}
\end{equation}
where $\Np(M,A)$ is the number of peaks with peak mass $\geq M$ in the survey area $A$. This shear peak abundance serves as our third cosmological probe.

In a weak-lensing survey, shear peaks can be readily identified and their amplitudes estimated using the observed image ellipticities of galaxies \citep[see, e.g.,][for an early work]{MiyazakiEtal2002}. Unfortunately, to date, there is no satisfactory method to predict the abundance of shear peaks for a given cosmology and survey design other than numerical simulations of lensing, which we employ here.

\subsection{Estimation of cosmological parameters}
\label{sec:theory:parameter_estimation}

We discuss the ability of future weak lensing surveys to constrain cosmological parameters within a Bayesian framework.
We consider a collection of model universes $\{\Model(\vect{\pi})\}$ characterized by a set of $\Nparam$ cosmological parameters $\vect{\pi}=\transposed{(\pi_1,\ldots,\pi_{\Nparam})}$, together with a set $\{\vect{d}\}$ of $\Ndata$-dimensional vectors $\vect{d}=\transposed{(d_1,\ldots,d_{\Ndata})}$ of observational data taken in order to constrain the parameters $\vect{\pi}$.
The knowledge about the probability of having the model universe with parameters $\vect{\pi}$ realized regardless of the outcome of the observations is assumed to be quantified by a \emph{prior} probability $\prior(\vect{\pi})$.
We also assume that we know the probability that the observations yield the data $\vect{d}$, given that the underlying model $\Model$ has parameters $\vect{\pi}$.
This probability is denoted $\likelihood(\vect{\pi}|\vect{d}) = p(\vect{d}|\vect{\pi})$, and referred to as the \emph{likelihood}.
The \emph{posterior} probability $\posterior(\vect{\pi}|\vect{d})$, i.e. the probability of having the parameters $\vect{\pi}$ realized in case the observations have yielded the data $\vect{d}$, can then be computed by:
\begin{equation}
  \posterior(\vect{\pi}|\vect{d}) = 
   \frac{\likelihood(\vect{\pi}|\vect{d}) \prior(\vect{\pi})}{\int\diff[N_\text{p}]{\vect{\pi}'}\, \likelihood(\vect{\pi}'|\vect{d}) p(\vect{\pi}') }
.
\end{equation}

\subsubsection{The likelihood}
\label{sec:theory:parameter_estimation:likelihood}

For simplicity, we approximate all relevant probabilities by piecewise constant or multivariate normal distributions. We assume a simple multivariate normal likelihood
\begin{equation}
\label{eq:likelihood_gaussian_data_approximation}
  \likelihood(\vect{\pi}|\vect{d}) =
  \frac{\left\lvert\matrb{C}_\text{d}^{-1}\right\rvert^{1/2}}{(2\pi)^{\Ndata/2}}
  \exp\left\{-\frac{1}{2}\transposed{\left[\vect{d} - \vect{\mu}(\vect{\pi}) \right]} \matrb{C}_\text{d}^{-1} \left[\vect{d} - \vect{\mu}(\vect{\pi}) \right] \right\},
\end{equation}
where the mean is given by a model prediction $\vect{\mu}(\vect{\pi})$, and the covariance matrix $\matrb{C}_\text{d}$ is assumed model-independent.
From a large set of $\Nreal$ ($\Nreal \gg 1$) independent realizations $\vect{d}_{r}$ of the data vector for a model with parameters $\vect{\pi}$, one can estimate the model prediction $\vect{\mu}$ by the sample mean $\est{\vect{\mu}}$,
\begin{equation}
\label{eq:df_sample_mean}
 \vect{\mu}(\vect{\pi}) \approx \est{\vect{\mu}} = \frac{1}{\Nreal} \sum_{r=1}^{\Nreal}\vect{d}_{r}
,
\end{equation}
and the data covariance $\matrb{C}_\text{d}$ by the sample covariance $\est{\matrb{C}}_\text{d}$,
\begin{equation}
\label{eq:df_sample_covariance}
\matrb{C}_\text{d} \approx \est{\matrb{C}}_\text{d} =
    \frac{1}{\Nreal - 1}
    \sum_{r=1}^{\Nreal}
               \bigl[\vect{d}_{r} - \vect{\mu}(\vect{\pi})\bigr]
   \transposed{\bigl[\vect{d}_{r} - \vect{\mu}(\vect{\pi})\bigr]}
.
\end{equation}
For normal distributions, the inverse covariance $\matrb{C}_\text{d}^{-1}$ is best estimated by \citep[][]{Anderson2003_book,HartlapEtal2007}:
\begin{equation}
\label{eq:df_sample_inverse_covariance}
\matrb{C}_\text{d}^{-1} \approx   
  \frac{\Nreal - \Ndata - 2}{\Nreal - 1} \bigl( \est{\matrb{C}}_\text{d} \bigr)^{-1}
,
\end{equation}
which requires $\Nreal > \Ndata + 2$ realizations.

The maximum-likelihood parameter set $\vect{\pi}_\text{ml}(\vect{d})$ for the measured data $\vect{d}$ is given by the equation
\begin{equation}
  \vect{0} = \transposed{\left.\parder{\vect{\mu}(\vect{\pi})}{\vect{\pi}}\right|_{\vect{\pi}_\text{ml}}} \matrb{C}_\text{d}^{-1} \left[\vect{d} - \vect{\mu}(\vect{\pi}_\text{ml})\right]
  .
\end{equation}
A Taylor expansion of the log-likelihood $\loglikelihood(\vect{\pi}|\vect{d})$ around the maximum-likelihood parameters $\vect{\pi}_\text{ml}$ up to second order in $\vect{\pi}$ yields the approximation
\begin{equation}
 \label{eq:likelihood_gaussian_parameter_approximation}
   \likelihood(\vect{\pi}|\vect{d}) =
     \frac{\left\lvert\matrb{C}_{\text{ml}}^{-1}\right\rvert^{1/2}}{(2\pi)^{\Nparam/2} }
     \exp \left[
   -\frac{1}{2}\transposed{\left(\vect{\pi} - \vect{\pi}_\text{ml}\right)} \matrb{C}_{\text{ml}}^{-1}  \left(\vect{\pi} - \vect{\pi}_\text{ml}\right)
  \right]
   ,
\end{equation}
where
\begin{equation}
   \matrb{C}_{\text{ml}}^{-1}(\vect{d}) =
  \begin{pmatrix}
   \displaystyle\sum_{k,l=1}^{\Ndata}
    \matrb{C}^{-1}_{\text{d},kl}
    \left[
     \dparder{\mu_k}{\pi_i} \dparder{\mu_l}{\pi_j}
     +  (d_k - \mu_k) \dparder{^2 \mu_l}{\pi_i \partial \pi_j}
    \right]_{\vect{\pi}_{\text{ml}}(\vect{d})}
   \end{pmatrix}_{i,j = 1}^{N_\text{p}}
  \!\!\!\!,
\end{equation}
is the inverse of the (maximum-likelihood approximation of the) parameter covariance matrix of the likelihood.

Eq.~\eqref{eq:likelihood_gaussian_parameter_approximation} reveals that the knowledge to be gained about the parameters $\vect{\pi}$ from the observations of a particular realization of the data $\vect{d}$ is encoded in the parameter covariance $\matrb{C}_{\text{ml}}(\vect{d})$ of the likelihood. The general information content of the likelihood can be quantified by the likelihood Fisher matrix, whose computation involves an ensemble average $\EV{\dots}_{\vect{d}}$ over realizations of the data:
\begin{equation}
  \matrb{F}_{\text{ml}} =
  \EV{\left(\parder{\ln L}{\pi_i}\parder{\ln L}{\pi_j}\right)_{i,j=1}^{\Nparam}}_{\vect{d}} 
  .
\end{equation}

In our study, where we assume that the likelihood is a multivariate normal distribution with constant parameter covariance and model predictions $\vect{\mu}$ that are well approximated by affine linear functions of the parameters $\vect{\pi}$, the likelihood Fisher matrix can be identified with the inverse of the likelihood parameter covariance:
\begin{equation}
\label{eq:simple_likelihood_parameter_covariance}
  \matrb{F}_{\text{ml}}
    = \matrb{C}_{\text{ml}}^{-1}
    = \transposed{\parder{\vect{\mu}}{\vect{\pi}}} \matrb{C}^{-1}_{\text{d}}  \parder{\vect{\mu}}{\vect{\pi}}
.
\end{equation}

Ignoring all correlations (i.e. approximating $\matrb{C}_{\text{ml}}$ and $\matrb{C}_{\text{d}}$ by diagonal matrices), Eq.~\ref{eq:simple_likelihood_parameter_covariance} can be recast into:
\begin{equation}
\label{eq:simple_likelihood_parameter_errors}
\left( \frac{1}{\sigma_{\text{ml},i}}\right)^{2} \approx \sum_{j=1}^{\Ndata} \left(\frac{1}{\sigma_{\text{d},j}} \totder{\mu_j}{\pi_i}\right)^{2},
\end{equation}
where $\sigma_{\text{ml},i} = \sqrt{\matr{C}_{\text{ml},ii}}$ denotes the likelihood standard deviation of an element of the parameter vector $\vect{\pi}$, and $\sigma_{\text{d},j} = \sqrt{\matr{C}_{\text{d},jj}}$ denotes the likelihood standard deviation of an element of the data vector $\vect{d}$.
Thus, each term $\bigl(\sigma_{\text{d},j}^{-1} \ttotder{\mu_j}{\pi_i}\bigr)$ of the sum in Eq.~\eqref{eq:simple_likelihood_parameter_errors} provides a rough measure of how much each datum $d_j$ contributes to the constraints on the parameter $\pi_i$.

\subsubsection{Prior and posterior distributions}
\label{sec:theory:parameter_estimation:priors_and_posteriors}

As one case, we consider a flat prior $p(\vect{\pi})$, i.e. one that is constant on some sufficiently large subset $\AllowedRegion \subseteq \R^{\Nparam}$ containing $\vect{\pi}_\text{ml}$. The posterior is then a truncated multivariate normal
\begin{equation}
\label{eq:posterior_for_flat_prior}
  \posterior(\vect{\pi}|\vect{d}) = 
  \begin{cases}
    c \exp \biggl[
   -\dfrac{1}{2}\transposed{\left(\vect{\pi} - \vect{\pi}_\text{po}\right)} \matrb{C}^{-1}_{\text{po}}  \left(\vect{\pi} - \vect{\pi}_\text{po}\right)
  \biggr]
    , & \vect{\pi} \in \AllowedRegion,  \\
    0, & \vect{\pi} \not\in \AllowedRegion, 
  \end{cases}
\end{equation} 
where $c$ is a normalization constant, the maximum of the posterior $\vect{\pi}_\text{po}(\vect{d}) = \vect{\pi}_\text{ml}(\vect{d})$, and the posterior parameter covariance matrix $\matrb{C}_{\text{po}}(\vect{d}) = \matrb{C}_{\text{ml}}(\vect{d})$.

As another case, we consider a Gaussian prior with mean $\vect{\pi}_{\text{pr}}$ and covariance $\matrb{C}_{\text{pr}}$. Then, the posterior is also a multivariate Gaussian
\begin{equation}
\label{eq:posterior_for_Gaussian_prior}
  \posterior(\vect{\pi}|\vect{d}) = 
   \frac{\left\lvert\matrb{C}_{\text{po}}^{-1}\right\rvert^{1/2}}{(2\pi)^{\Nparam/2} }
     \exp \left[
   -\dfrac{1}{2}\transposed{\left(\vect{\pi} - \vect{\pi}_\text{po}\right)} \matrb{C}^{-1}_{\text{po}}  \left(\vect{\pi} - \vect{\pi}_\text{po}\right)
  \right]
\end{equation} 
with mean
\begin{equation}
  \vect{\pi}_{\text{po}} (\vect{d}) =  \left[\matrb{C}_{\text{pr}}^{-1} +   \matrb{C}_{\text{ml}}^{-1}  (\vect{d}) \right]^{-1} 
  \left[\matrb{C}_{\text{pr}}^{-1}\vect{\pi}_\text{pr} +   \matrb{C}_{\text{ml}}^{-1} (\vect{d}) \vect{\pi}_\text{ml}(\vect{d}) \right]
,
\end{equation} 
and covariance matrix
\begin{equation}
  \matrb{C}_{\text{po}} (\vect{d}) = \left[\matrb{C}_{\text{pr}}^{-1} + \matrb{C}_{\text{ml}}^{-1}  (\vect{d}) \right]^{-1}
.
\end{equation}

\subsubsection{Conditional and marginal posterior distributions}
\label{sec:theory:parameter_estimation:conditional_and_marginal_distributions}

For a Gaussian posterior distribution, conditional and marginal distributions are also Gaussian. Consider a partition of the parameters $\vect{\pi}=(\vect{\pi}_\text{i},\vect{\pi}_\text{u})$ into $N_\text{p,i}$ `interesting' parameters $\vect{\pi}_\text{i}$ and $N_\text{p,u} = \Nparam - N_\text{p,i}$ `uninteresting' parameters $\vect{\pi}_\text{u}$. Similarly, the posterior maximum is split into $\vect{\pi}_{\text{po}} = (\vect{\pi}_{\text{po,i}}, \vect{\pi}_{\text{po,u}})$. 
The posterior covariance matrix and its inverse are split into submatrices,
\begin{equation}
\matrb{C}_{\text{po}}
=
\begin{pmatrix}
  \matrb{C}_{\text{po,ii}} & \matrb{C}_{\text{po,iu}} \\
  \matrb{C}_{\text{po,ui}} & \matrb{C}_{\text{po,uu}}
\end{pmatrix}
\quad\text{and}\quad    
\matrb{C}_{\text{po}}^{-1}
=
\begin{pmatrix}
  \bigl(\matrb{C}_{\text{po}}^{-1}\bigr)_\text{ii} & \bigl(\matrb{C}_{\text{po}}^{-1}\bigr)_\text{iu} \\
  \bigl(\matrb{C}_{\text{po}}^{-1}\bigr)_\text{ui} & \bigl(\matrb{C}_{\text{po}}^{-1}\bigr)_\text{uu}
\end{pmatrix}
.
\end{equation}
Here, for example, $\matrb{C}_{\text{po,iu}}$ denotes the submatrix obtained from the covariance matrix $\matrb{C}_{\text{po}}$ by only keeping the rows of the `interesting' parameters and the columns of the `uninteresting' parameters, whereas $\bigl(\matrb{C}_{\text{po}}^{-1}\bigr)_\text{iu}$ denotes the submatrix obtained by only keeping the `interesting' rows and `uninteresting' columns of the inverse covariance matrix $\matrb{C}_{\text{po}}^{-1}$.

The marginal distribution of $\vect{\pi}_\text{i}$ is then given by
\begin{equation}
\label{eq:marginal_for_Gaussian_posterior}
  \posterior(\vect{\pi}_\text{i}|\vect{d}) = 
   \frac{\left\lvert\matrb{C}_{\text{po,ii}}^{-1}\right\rvert^{1/2}}{(2\pi)^{N_\text{p,i}/2} }
     \exp \left[
   -\dfrac{1}{2}\transposed{\left(\vect{\pi}_\text{i} - \vect{\pi}_\text{po,i}\right)} \matrb{C}^{-1}_{\text{po,ii}}  \left(\vect{\pi}_\text{i} - \vect{\pi}_\text{po,i}\right)
  \right]
.
\end{equation}
From this follows that the posterior marginal variance on a single parameter $\pi_i$ is given by $\matrb{C}_{\text{po},ii}$.

The conditional distribution  of $\vect{\pi}_\text{i}$ is given by 
\begin{equation}
\label{eq:conditional_for_Gaussian_posterior}
  \posterior(\vect{\pi}_\text{i}|\vect{d}, \vect{\pi}_\text{u}) = 
   \frac{\left\lvert \matrb{C}_{\text{c,ii}}^{-1} \right\rvert^{1/2} \!\!\!}{\! (2\pi)^{N_\text{p,i}/2} \!\!}
     \exp \left[
   -\dfrac{1}{2}\transposed{\left(\vect{\pi}_\text{i} - \vect{\pi}_\text{c,i}\right)} \matrb{C}_{\text{c,ii}}^{-1} \left(\vect{\pi}_\text{i} - \vect{\pi}_\text{c,i}\right)
  \right]
,\!
\end{equation}
where
\begin{equation}
\label{eq:conditional_mean_for_Gaussian_posterior}
\begin{split}
  \vect{\pi}_\text{c,i}(\vect{d}, \vect{\pi}_\text{u}) 
  &= 
  \vect{\pi}_{\text{po,i}} (\vect{d}) 
  -  \bigl[\bigl(\matrb{C}_{\text{po}}^{-1}\bigr)_\text{ii}\bigr]^{-1} \bigl(\matrb{C}_{\text{po}}^{-1}\bigr)_\text{iu} \left[ \vect{\pi}_\text{u} - \vect{\pi}_{\text{po,u}}(\vect{d}) \right]
  \\&= 
  \vect{\pi}_{\text{po,i}} (\vect{d}) 
  + \matrb{C}_{\text{po,iu}} \bigl(\matrb{C}_{\text{po,uu}}\bigr)^{-1} \left[ \vect{\pi}_\text{u} - \vect{\pi}_{\text{po,u}}(\vect{d}) \right]
,
\end{split}
\end{equation}
and
\begin{equation}
\label{eq:conditional_covariance_for_Gaussian_posterior}
\begin{split}
  \matrb{C}_{\text{c,ii}}^{-1}
   &= 
   \bigl(\matrb{C}_{\text{po}}^{-1}\bigr)_\text{ii} 
= 
\left[ \matrb{C}_{\text{po,ii}} -  \matrb{C}_{\text{po,iu}}  \bigl(\matrb{C}_{\text{po,uu}}\bigr)^{-1} \matrb{C}_{\text{po,ui}}  \right]^{-1}
 .
\end{split}
\end{equation}
From this follows that the posterior conditional variance on a single parameter $\pi_i$ is given by $1/\bigl(\matrb{C}^{-1}_{\text{po}}\bigr)_{ii}$.

Furthermore, fixing certain parameters to values differing from the unconditional posterior mean may shift the conditional posterior mean values of the remaining parameters to values different from their unconditional posterior mean values. Such shifts may be the cause of biases in a parameter estimation. If one or more model parameters are wrongly assumed to take values different from their true values, the estimates for the remaining parameters will also be `wrong' on average unless there is no correlation between the parameters.

\section{Methods}
\label{sec:methods}

We use $N$-body simulations of cosmic structure formation alongside with ray-tracing to create a large suite of mock weak-lensing observations in various CDM cosmologies. From these mock observations, we compute the expected signals for the cosmic-shear correlation functions, the aperture mass statistics, and the shear peak statistics. Furthermore, we estimate the covariance of these lensing statistics and their dependence on cosmological parameters. The results for the parameter dependence and covariance of the lensing statistics are then used to forecast parameter errors for future large weak-lensing surveys and discuss possible biases arising from neglecting non-Gaussianity.

\subsection{N-body simulations}
\label{sec:methods:N_body_simulations}

We use two sets of $N$-body simulations of structure formation in flat CDM cosmologies for our work. To study the dependence of weak lensing observables on primordial non-Gaussianity, we use the (non-)Gaussian simulations described in \citet{DesjacquesSeljakIliev2009} and \citet{MarianEtal2011}. These assume a flat $\Lambda$CDM cosmology whose parameters are compatible with WMAP 5-year data \citep{KomatsuEtal2009}:
$h = 0.7,$ $\Omega_\text{m} = 0.279$, $\Omega_\text{b} = 0.0462$, $\Omega_\text{DE} = 0.721$, $w_0 = -1$, $\sigma_8 = 0.81$, $n_\text{s} = 0.96$, and $\fNL \in \{-100,0,100\}$.
The simulations employ \softwarename{Gadget-2} \citep[][]{Springel2005_GADGET2} to follow $1024^3$ particles in a cubic volume of $1600 h^{-1}\,\Mpc$ side length with a force softening length of $40 h^{-1}\,\kpc$.

There are six (non-)Gaussian simulations for each value of $\fNL$. The initial conditions of the simulations are matched between the different $\fNL$ values, which reduces the influence of cosmic variance on the relative values of measured weak-lensing observations for different $\fNL$. We choose the six simulations with $\fNL = 0$ to represent our fiducial cosmology.

To study the dependence of weak lensing observables on a number of other parameters, we make use of the \softwarename{zHORIZON} simulations by \citet{Smith2009}.
These simulations use \softwarename{Gadget-2} with $750^3$ particles to follow the structure formation in a cubic region of $1500 h^{-1}\,\Mpc$ side length in flat CDM cosmologies with parameters around the best-fitting WMAP 5-year values.

There are eight \softwarename{zHORIZON} simulations for a flat CDM cosmology with $h = 0.7,$ $\Omega_\text{m} = 0.25$, $\Omega_\text{b} = 0.04$, $\Omega_\text{DE} = 0.75$, $w_0 = -1$, $\sigma_8 = 0.8$, $n_\text{s} = 1$, and $\fNL = 0$. Furthermore, there are 32 simulations that vary one of the above parameters, four each for $\Omega_\text{m} \in \{0.2,0.3\}$, $\sigma_8 \in \{0.7,0.9\}$, $n_\text{s} \in \{0.95, 1.05\}$, and $w_0 \in \{0.8, 1.2 \}$. Together with the (non-)Gaussian simulations, these allow us to study the change of observables with varying $\fNL$, $\Omega_\text{m}$, $\sigma_8$, $n_\text{s}$, or $w_0$.

\subsection{Lensing simulations}
\label{sec:methods:lensing_simulations}

For the lensing simulations, we closely follow the method of \citet{HilbertEtal2009_RT}, which employs a Multiple-Lens-Plane ray-tracing algorithm \citep[e.g.][]{BlandfordNarayan1986,JainSeljakWhite2000} to calculate the light deflection by matter structures in the simulations. The past lightcone of a fiducial observer in a simulation is partitioned into redshift slices with about $120 h^{-1}\,\Mpc$ comoving line-of-sight thickness (assuming the flat-sky approximation). Each slice is then filled with the matter structures of the simulation snapshot closest in redshift to the mean redshift of the slice, and the slice content is projected onto a lens plane. Light rays are then traced back from the observer through these lens planes, at which the light rays are deflected. The deflection angles (and their derivatives) at the lens planes are calculated from the projected matter distribution by a mesh method with a resolution of $20 h^{-1}\,\kpc$ comoving.

For each simulation, we generate 16 (quasi-)independent fields of view by choosing different observer positions within the simulation box. Each field has an area of $12\times12\,\degt^2$ and is covered by a regular mesh of $4096^2$ pixels in the image plane, which provides an angular resolution of $10\,\arcsect$. Thus we obtain 96 simulated survey fields covering a total area of $14\,000\,\degt^2$ for each $\fNL$-value from the (non-)Gaussian simulations. Furthermore, we have 128 fields and a total area of $18\,000\,\degt^2$ for the fiducial cosmology of the \softwarename{zHORIZON} simulations, and $9\,000\,\degt^2$ for the other cosmologies of the \softwarename{zHORIZON} simulations.

For each pixel in the simulated fields, a light ray is traced back through the planes up to source redshift $z=4$. The ray distortions along the rays are then used to calculate the effective convergence and shear for a population of sources with median redshift $\zmedian=0.9$ and redshift distribution \citep[][]{BrainerdBlandfordSmail1996}
\begin{equation}
\label{eq:source_redshift_distribution}
 p_{z}(z)=\frac{3z^2}{2z_0^3}\exp\left[-\left(\frac{z}{z_0}\right)^{3/2} \right],
 \text{ where }
 z_0 = 0.71 \zmedian.
\end{equation}
This redshift distribution is similar to those expected for the planned lensing surveys with \satellitename{Euclid} \citep{LaureijsEtal2011_Euclid_DSR} or the \telescopename{LSST} \citep[][]{LSST_Science_Book_2009}.

The resulting simulated shear and convergence maps are free of noise due to the finite number of source galaxies and their intrinsic ellipticities. For a realistic description of the measurement uncertainties, we also need to take this shape noise into account. We thus create shear maps with shape noise by adding Gaussian noise with standard deviation $\sigmapixel = \sigmaepsilongalpc/\sqrt{\ngal \Apixel}$ to each pixel (with pixel area $\Apixel \approx 10^2\,\arcsect^2$) of the simulated shear fields. We assume an intrinsic galaxy ellipticity distribution with standard deviation $\sigmaepsilongalpc = 0.3$ for each ellipticity component, and a source galaxy density $\ngal = 40\,\arcmint^{-2}$.

\subsection{Estimation of the model predictions}
\label{sec:methods:model_predictions}

We use the simulated lensing fields to compute the expected signals for the cosmic-shear correlation functions $\xi_\pm(\vartheta)$ in bins of separation $\vartheta$, the third-order aperture mass statistics $\bEV{\Map^3}(\vartheta)$ as a function of filter scale $\vartheta$, and the shear peak abundance $\dpa$ in bins of filter mass $M$.

For each field, we estimate the shear correlation functions from the shape-noise free shear maps employing the Fast Fourier Transform (FFT) technique described in \citet{HilbertHartlapSchneider2011}. The two-dimensional auto- and cross-correlations of the shear components $\gamma_1$ and $\gamma_2$ are computed employing FFTs. The resulting two-dimensional correlations $\bEV{\gamma_i \gamma_j}(\vvartheta)$ are then multiplied by direction-dependent phase factors depending on the direction of $\vvartheta$ and binned according to separation $|\vvartheta|$ to obtain the binned cosmic shear correlations.

Estimates for the aperture mass statistics are computed from the shape-noise free convergence maps of the fields. The convergence maps are convolved with the filter \eqref{eq:M_ap_filter_function_U} of a given scale $\vartheta$ by using FFTs and exploiting the convolution theorem. The prediction for the aperture statistics $\bEV{\Map^3}(\vartheta)$ is then computed by a simple average of the cubed values of the filtered convergence field $\Map(\vtheta,\vartheta)$, whereby points $\vtheta$ closer than $4\vartheta$ to a field boundary are discarded. The procedure is carried out for the different filter scales $\vartheta$.

Estimates for the shear peak abundance are obtained from the noisy shear maps using the hierarchical algorithm described in Section \ref{sec:theory:weak_lensing_observables:shear_peak_abundance}.\footnote{
Unlike the predictions for the shear correlations or the aperture statistics, the model predictions for the peak abundance depend on the level of shape noise.
}
We employ a series of 12 filters logarithmically spanning the range $M = 8.85 \times 10^{13}h^{-1}\,\Msolar$ to $2\times 10^{15}h^{-1}\,\Msolar$, with a signal-to-noise detection threshold $(S/N)_{\mathrm{min}}=2.6$. Linear interpolation is used to estimate the filter mass $M$ with the best matching amplitude for each peak. Finally, peaks too close to larger peaks (i.e. within the projected virial radius of an NFW halo at the model redshift with the same amplitude as the larger peak) are discarded, and the remaining peaks are binned according to their peak mass $M$. Note that we do not use any information about any actual halos that create some of the peaks.

For each cosmology, we calculate model predictions for the considered lensing statistics by a simple average of the statistics estimated in each of the 64 to 128 simulated fields of that cosmology according to Eq.~\eqref{eq:df_sample_mean}. The model predictions for the different lensing statistics and scales are then combined into one vector $\vect{\mu}=\bigl(\xi_+(\vartheta_1),\xi_+(\vartheta_2),\ldots,\xi_-(\vartheta_1),\ldots,\bEV{\Map^3}(\vartheta_1),\ldots,\dpa(M_1),\ldots \bigr)$.

We assume for simplicity that the model predictions $\vect{\mu}$ are linear functions of the cosmological parameters $\vect{\pi}=(\Omega_\text{m}, \sigma_8, n_\text{s}, w_0, \fNL)$ in the range covered by our simulations. We compute the derivatives of the model predictions w.r.t. the cosmological parameters by a central difference:
\begin{equation}
\label{eq:finite_differences_for_predictions_and_parameters}
\parder{\mu_i}{\pi_j} = 
  \frac
  {\mu_i\bigl(\vect{\pi}^{(j,+1)}\bigr) - \mu_i\bigl(\vect{\pi}^{(j,-1)}\bigr)}
  { \pi^{(j,+1)}_j - \pi^{(j,-1)}_j}
.
\end{equation}
Here, $\vect{\pi}^{(j,+1)}$ and $\vect{\pi}^{(j,-1)}$ are two sets of parameters that differ only in the $j$-th component. Our particular choices of parameters for the simulations (see Sec.~\ref{sec:methods:N_body_simulations}) ensures that we have predictions for two suitable sets of parameters available for all partial derivatives. We also check that using forward or backward differences instead central differences does not yield significant differences in the estimated derivatives.

\subsection{Estimation of the measurement uncertainties and data covariances}
\label{sec:methods:covariances}

We estimate the covariance of the lensing statistics from the sample covariance of the measured statistics in the 96 $12\times12\,\deg^2$ simulated fields created from our Gaussian simulations.\footnote{
As a check, we also compute the covariances from the simulated fields of the non-Gaussian simulations, and from the fields of the fiducial model of the  \softwarename{zHORIZON} simulations. The resulting covariances are very similar to those estimated from the Gaussian simulations.
}
To obtain realistic covariance estimates, we measure the shear correlation functions, the aperture mass statistics, and the shear peak functions from the shear maps with shape noise.

For the noisy measurements of the shear correlations functions $\xi_\pm$, we use the same method as for the model predictions, but with the noisy shear maps as input. The noisy aperture mass maps for measuring $\bEV{\Map^3}$ are obtained by filtering the noisy complex shear maps with the filter \eqref{eq:M_ap_filter_function_Q} modified by a phase factor to select the tangential shear component. The shear peak counts are obtained from the same noisy shear maps using the hierarchical filtering. The resulting measurements of $\xi_\pm$, $\bEV{\Map^3}$, and $\dpa$ in bins of separation or scale obtained from each noisy shear field $r$ are arranged into a noisy data vector $\vect{d}_r=\transposed{\bigl(\xi_+(\vartheta_1)_r,\ldots\bigr)}$.

The data covariance matrices and inverse data covariance matrices are calculated from the sample covariance of the noisy measurements in each simulated field by using Eqs.~\eqref{eq:df_sample_covariance} and \eqref{eq:df_sample_inverse_covariance}. The resulting matrices describe the expected data covariance for one field of $144\,\deg^2$. We rescale the matrices to describe covariances of a survey with 125 independent $144\,\deg^2$ fields, i.e. with a total area of $18\,000\,\degt^2$. With that area, a median source redshift $\zmedian = 0.9$, a source galaxy density $\ngal = 40\,\arcmint^{-2}$, and a standard deviation $\sigmaepsilongalpc = 0.3$ per ellipticity component, the resulting covariances are representative of a large future `all-sky' weak lensing survey similar to those planned for \satellitename{Euclid} or the LSST.

The covariances calculated in that way probably underestimate the statistical errors for such large surveys. Since our simulated fields are created from only six independent simulations and therefore are partially correlated, the sample covariance of the fields might underestimate the true covariance expected for such fields. Furthermore, due to large-scale cosmic variance, the data measured in a survey covering half the sky is more correlated than data from a hypothetical survey with the same total area but distributed in many independent fields.

We are also interested in the statistical uncertainty of our model predictions and the relative importance of shape noise and cosmic variance. If only shape noise and cosmic variance are considered as sources of noise, the covariance in the measured shear correlation functions and aperture statistics can be split into several parts: a pure shape noise part, a pure cosmic variance part, and mixed noise parts due to a coupling between shape noise and shear correlations \citep[e.g.][]{SchneiderEtal2002,KilbingerSchneider2005}. We estimate the cosmic variance contribution to the cosmic shear correlation functions and the third-order aperture statistics from the sample covariance of the shape noise-free simulations. The expected pure shape noise and mixed noise contributions are computed from the assumed source galaxy density, intrinsic ellipticity variance, and measured shear correlations.

\section{Results}
\label{sec:results}

In the following, we discuss the effects of primordial non-Gaussianity of the local type on weak lensing observables, and how these effects can be used in large surveys to constrain the non-Gaussianity parameter $\fNL$ and other cosmological parameters. In particular, we present the separate and joint parameter constraints obtained from cosmic shear correlations, third-order aperture statistics, and shear peak statistics. Furthermore, we discuss the biases in other cosmological parameters induced when a non-zero non-Gaussianity is neglected in the parameter inference from weak lensing data.

As indicated in the outline of our simulations, we discuss results expected from a Euclid- or LSST-like lensing survey with an area of $18\,000\,\degt^2$, a source galaxy population with a median redshift $\zmedian = 0.9$, a number density $\ngal = 40\,\arcmint^{-2}$, and a standard deviation $\sigmaepsilongalpc = 0.3$ per intrinsic ellipticity component. We assume the correlation functions $\xi_\pm(\vartheta)$ are measured in 10 logarithmically spaced bins in the range $0.5\,\arcmint \leq \vartheta \leq 120\,\arcmint$. The aperture mass statistics $\bEV{\Map^3}(\vartheta)$ is measured for 10 logarithmically spaced filter scales $\vartheta$ in the range $0.5\,\arcmint \leq \vartheta \leq 60\,\arcmint$. The peak abundance $\dpa(M)$ is measured in 10 logarithmically spaced bins in the range $10^{14}\,\Msolar \leq M \leq 1.3\times10^{15}\,\Msolar$.

We only consider uncertainties in the measurements due to shape noise (i.e. due to the unknown intrinsic shapes of individual galaxies) and cosmic variance (i.e. due to the variation in structures between particular portions of the Universe), which set a fundamental limit on the accuracy of the results from a weak lensing survey. We do not account for other contribution to the error budget (e.g. from uncertainties in the model predictions or biases related to imperfect galaxy shape and redshift measurements in observations).

\subsection{Effects of primordial non-Gaussianity on lensing observables}
\label{sec:results:observables}

First, we study how primordial non-Gaussianity changes the observed shear correlation functions, aperture statistics, and peak statistics compared to the Gaussian case. The changes in the lensing statistics due to non-Gaussianity are usually small compared to the dynamic range in which the signals vary in the considered range of angular scales or masses. We thus discuss relative differences in the signals for different $\fNL$.


\begin{figure}
\centerline{\includegraphics[width=\linewidth]{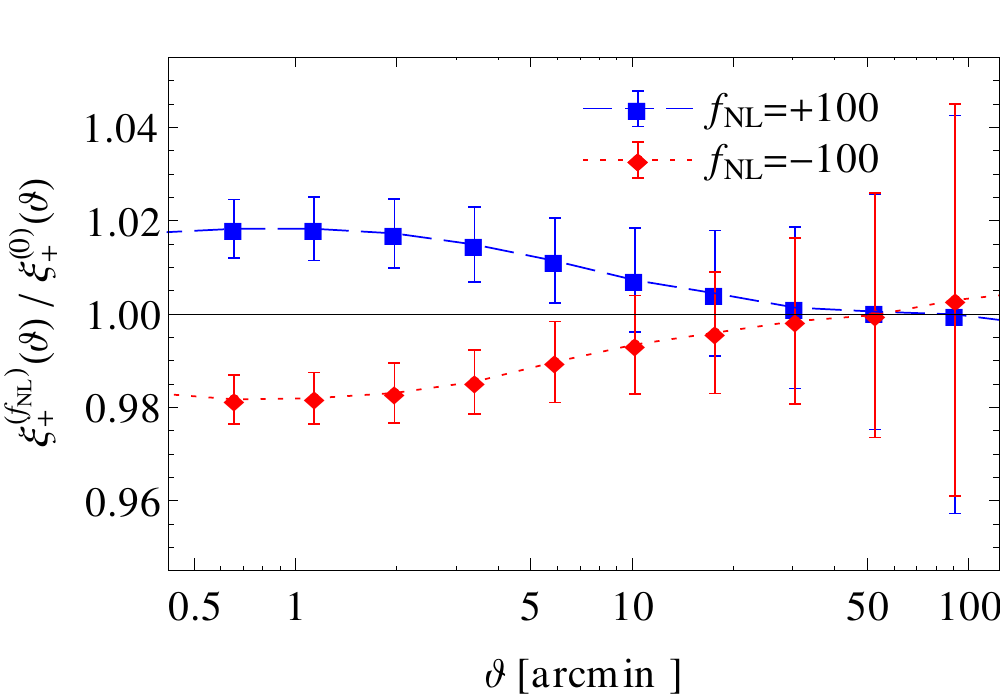}}
\centerline{\includegraphics[width=\linewidth]{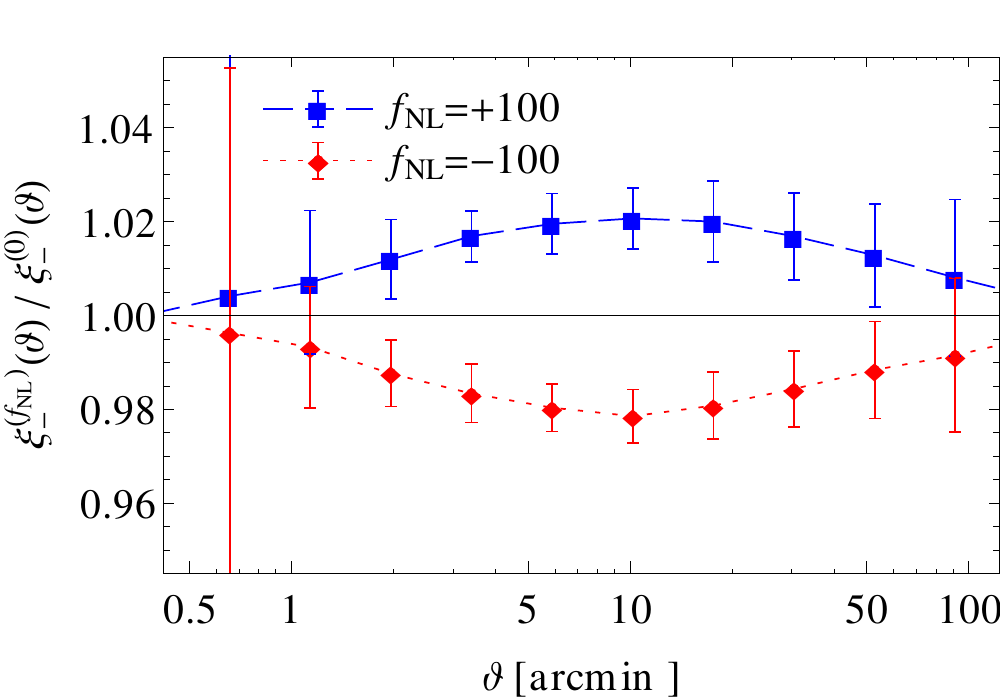}}
\caption{
\label{fig:xi_pm_ratio_for_f_nl}
Cosmic shear correlations $\xi_\pm^{(\fNL)}(\vartheta)$ as functions of separation $\vartheta$ for different values of non-Gaussianity $\fNL$. Points show the relative difference between $\fNL \neq 0$ and $\fNL = 0$. Error bars indicate the statistical uncertainties due to shape noise and cosmic variance expected for large future surveys.
}
\end{figure}

Figure~\ref{fig:xi_pm_ratio_for_f_nl} shows the relative differences in the cosmic shear correlation functions $\xi_+$ and $\xi_-$ between the Gaussian case $\fNL=0$ and the non-Gaussian case $\fNL\neq 0$. The error bars indicate the statistical errors due to the intrinsic galaxy image shapes and cosmic variance expected for a large future survey (see Section~\ref{sec:results:error_analysis} for a more detailed discussion).

On scales $\vartheta < 1 \,\degt$, the shear correlation function $\xi_+(\vartheta)$ is larger for larger $\fNL$. The largest relative difference is visible on scales $\vartheta \sim 1 \,\arcmint$. The difference is only about $\pm2\%$ for $\fNL = \pm 100$, but this is large enough to be detectable in future surveys, since the errors are $\lesssim 1\%$ at these separations, as shown in the next section. The results for $\xi_-$ are very similar, but with the largest relative differences shifted to larger scales $\sim 10\,\arcmint$.

Both $\xi_+$ and $\xi_-$ are weighted projections of the three-dimensional matter correlation, with $\xi_+(\vartheta)$ probing larger 3-D scales than $\xi_-(\vartheta)$ at the same angular scale $\vartheta$. At redshifts $z\sim0.3$, where lensing is most efficient, the angular scales where the shear correlations show the largest response to $\fNL$ correspond to 3-D scales of a few hundred $\kpc$.
This result is consistent with predictions from a halo model \citep{SmithDesjacquesMarian2011}. In the presence of primordial non-Gaussianity, both the abundance and density profiles of massive haloes are slightly increased/decreased for positive/negative $\fNL$. The relative contribution of these massive halos to the matter correlation is largest at the group and cluster scales. On smaller length scales, the signal becomes dominated by the lower mass halos, for which the effects of primordial non-Gaussianity are significantly reduced. On larger scales, two-halo terms, which are also less affected by primordial non-Gaussianity, dominate the matter correlation. This signature is qualitatively mirrored in Figure~\ref{fig:xi_pm_ratio_for_f_nl}.

\begin{figure}
\centerline{\includegraphics[width=\linewidth]{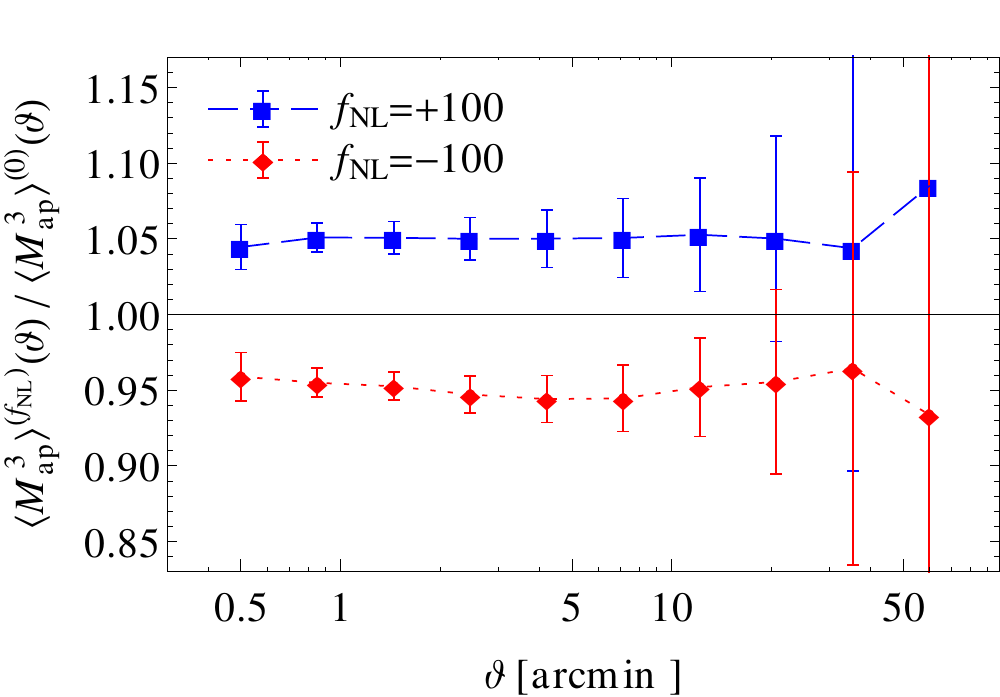}}
\caption{
\label{fig:M_ap_3_ratio_for_f_nl}
Third-order aperture mass statistics $\bEV{\Map^3}^{(\fNL)}(\vartheta)$ as function of filter scale $\vartheta$ for different values of non-Gaussianity $\fNL$. Points show the relative difference between $\fNL \neq 0$ and $\fNL = 0$. Error bars on indicate the statistical uncertainties due to shape noise and cosmic variance expected for large future surveys.
}
\end{figure}

Figure~\ref{fig:M_ap_3_ratio_for_f_nl} shows the differences in the aperture statistics $\bEV{\Map^3}(\vartheta)$. Similar to the shear correlation functions, the third-order aperture mass statistics is larger for larger primordial non-Gaussianity. The relative difference in $\bEV{\Map^3}(\vartheta)$ is about $\pm5\%$ for $\fNL\pm 100$, which is clearly larger than the statistical uncertainties of about $1\%$ expected for scales $\vartheta \lesssim 5\,\arcmint$ in large future surveys, as shown in the next section.

\begin{figure}
\centerline{\includegraphics[width=\linewidth]{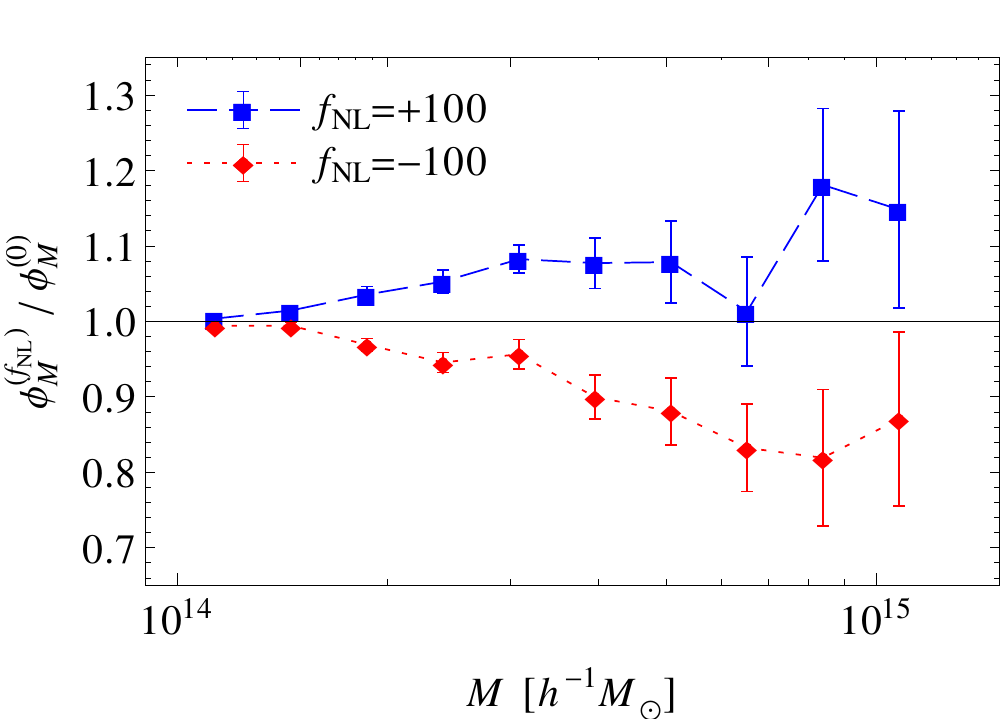}}
\caption{
\label{fig:peak_abundance_ratio_for_f_nl}
Shear peak abundance $\dpa^{(\fNL)}(M)$ as function of peak mass $M$ for different values of non-Gaussianity $\fNL$. Points show the relative difference between $\fNL \neq 0$ and $\fNL = 0$. Error bars indicate the statistical uncertainties due to shape noise and cosmic variance expected for large future surveys.
}
\end{figure}

The impact of primordial non-Gaussianity on the shear peak abundance $\dpa(M)$ is shown in Fig.~\ref{fig:peak_abundance_ratio_for_f_nl}. The peak abundance increases with increasing $\fNL$. Similar results were presented in our previous study \citep{MarianEtal2011}. The change is stronger for larger peak masses $M$. For these larger masses, however, also the noise becomes larger.

\subsection{Error analysis}
\label{sec:results:error_analysis}

\begin{figure}
\centerline{\includegraphics[width=\linewidth]{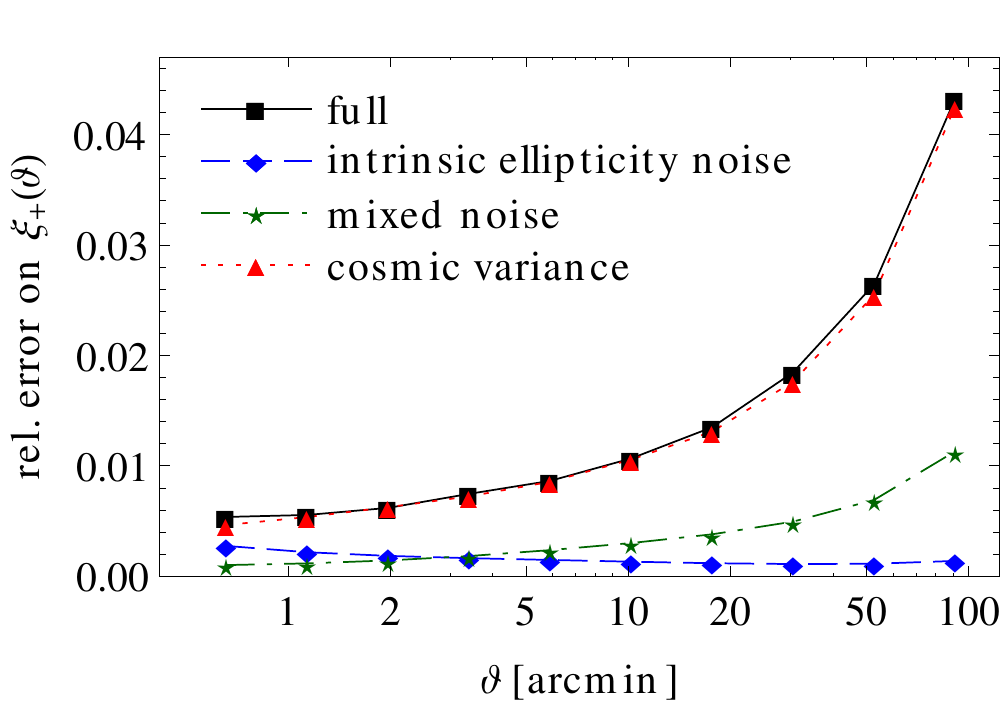}}
\centerline{\includegraphics[width=\linewidth]{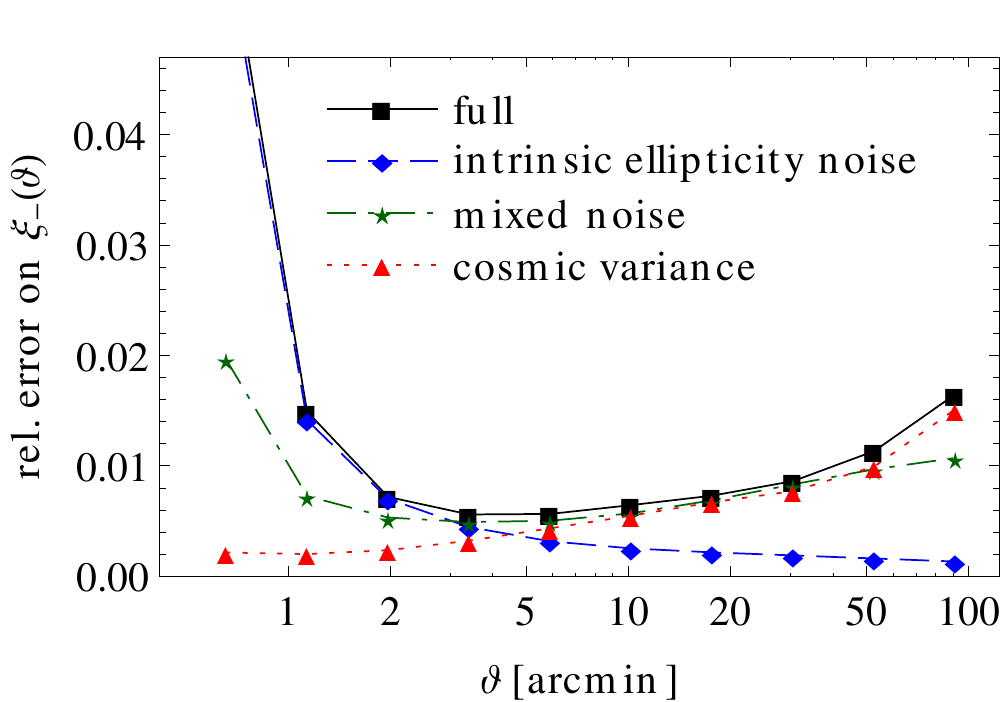}}
\caption{
\label{fig:xi_pm_relative_error}
Expected contributions to the relative errors (symbols, connected by lines for visual guidance) on the estimates of the shear correlation functions $\xi_{\pm}(\vartheta)$ in bins of separation $\vartheta$: full error (squares), intrinsic ellipticity noise (diamonds), mixed noise (stars), and cosmic variance (triangles).
}
\end{figure}

As discussed in Section~\ref{sec:methods:covariances}, the uncertainties in the measured shear correlation functions can be split into a pure shape noise part, a pure cosmic variance part, and a mixed noise part due to a coupling of shape noise and shear correlations. In Fig.~\ref{fig:xi_pm_relative_error}, these contributions are shown together with the full statistical uncertainties for our assumed survey. The cosmic variance dominates the error on $\xi_+$ on all considered scales \citep[cf., e.g., ][]{BeynonBaconKoyama2010}. The cosmic variance sets a lower limit of $\sim 0.5\%$ on the uncertainty in $\xi_+(\vartheta)$ for separations $\vartheta \sim 1\,\arcmint$, and of $\gtrsim 1\%$ on $\vartheta \gtrsim 10\,\arcmint$. Shape noise is the dominant contribution only for $\xi_-$ on scales $\vartheta \lesssim 2\,\arcmint$. On larger scales, the cosmic variance and mixed noise dominate the error on $\xi_-$.

As previously mentioned, our covariance results based on the simulated $12\times12\,\degt^2$ fields might underestimate the covariance of $\xi_\mp(\vartheta)$ in a full-sky survey, in particular for large separations $\vartheta$. To check this, we also compute the expected full-sky covariance employing the normal and the log-normal approximation \citep{SchneiderVanWaerbekeMellier2002,HilbertHartlapSchneider2011}. The results from these analytic approximations indicate that our simulation results do not underestimate the covariance except for $\xi_+$ on large scales $\vartheta\gtrsim 1\,\degt$, where the difference is $\sim 20\%$.

\begin{figure}
\centerline{\includegraphics[width=\linewidth]{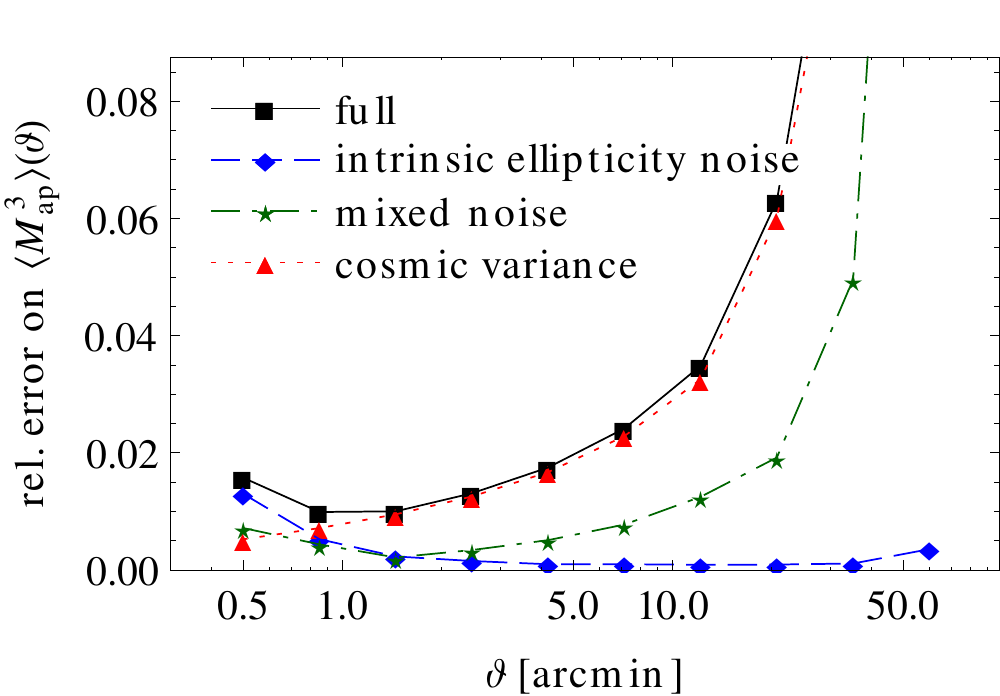}}
\caption{
\label{fig:M_ap_3_relative_error}
Expected contributions to the relative errors on the estimates of the third-order aperture mass statistics $\bEV{\Map^3}(\vartheta)$ for various filter scales $\vartheta$: full error (squares), intrinsic ellipticity noise (diamonds), mixed noise (stars), and cosmic variance (triangles).
}
\end{figure}

Figure~\ref{fig:M_ap_3_relative_error} shows that the cosmic variance is also the dominant contribution to the error on the aperture statistic $\bEV{\Map^3}(\vartheta)$ on scales $\vartheta > 1\,\arcmint$. Only for smaller filter scales, there is a significant contribution from galaxy shape noise and mixed noise.

\begin{figure}
\centerline{\includegraphics[width=\linewidth]{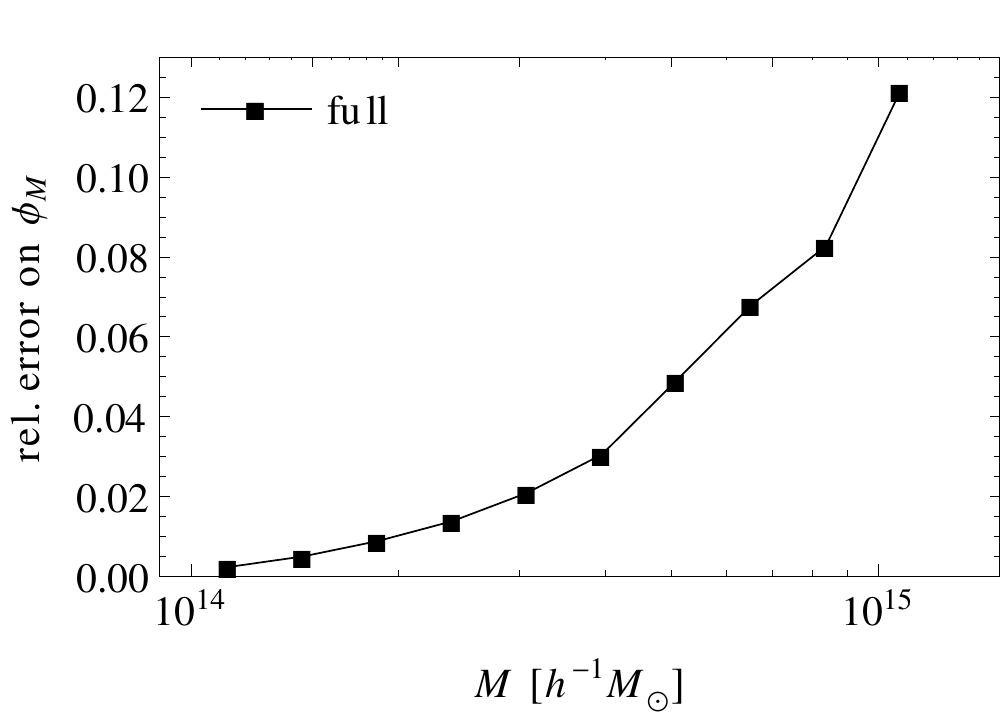}}
\caption{
\label{fig:peak_abundance_relative_error}
Expected relative errors (symbols, connected by lines for visual guidance) on the estimates of the shear peak abundance $\dpa(M)$ for various peak masses $M$.
}
\end{figure}

The estimated errors for the shear peak abundance are shown in Figure~\ref{fig:peak_abundance_relative_error}. The errors are $\sim 1\%$ for the bins with peak masses of a few $10^{14}h^{-1}\,\Msolar$ or smaller, but rapidly increase to tens of percent for larger peak masses. There is no easy way to split the errors into contributions from shape noise or cosmic variance. However, by comparing results from fields with the same cosmic structure but different shape noise realizations, we find that shape noise makes a substantial contribution to the statistical errors on the peak counts even for intermediate and large peak masses. A more thorough understanding of the noise properties of the shear peak statistics is certainly worthwile in future.

\subsection{The covariance of the lensing statistics}
\label{sec:results:covariance}

\begin{figure*}
\centerline{\includegraphics[width=0.7\linewidth]{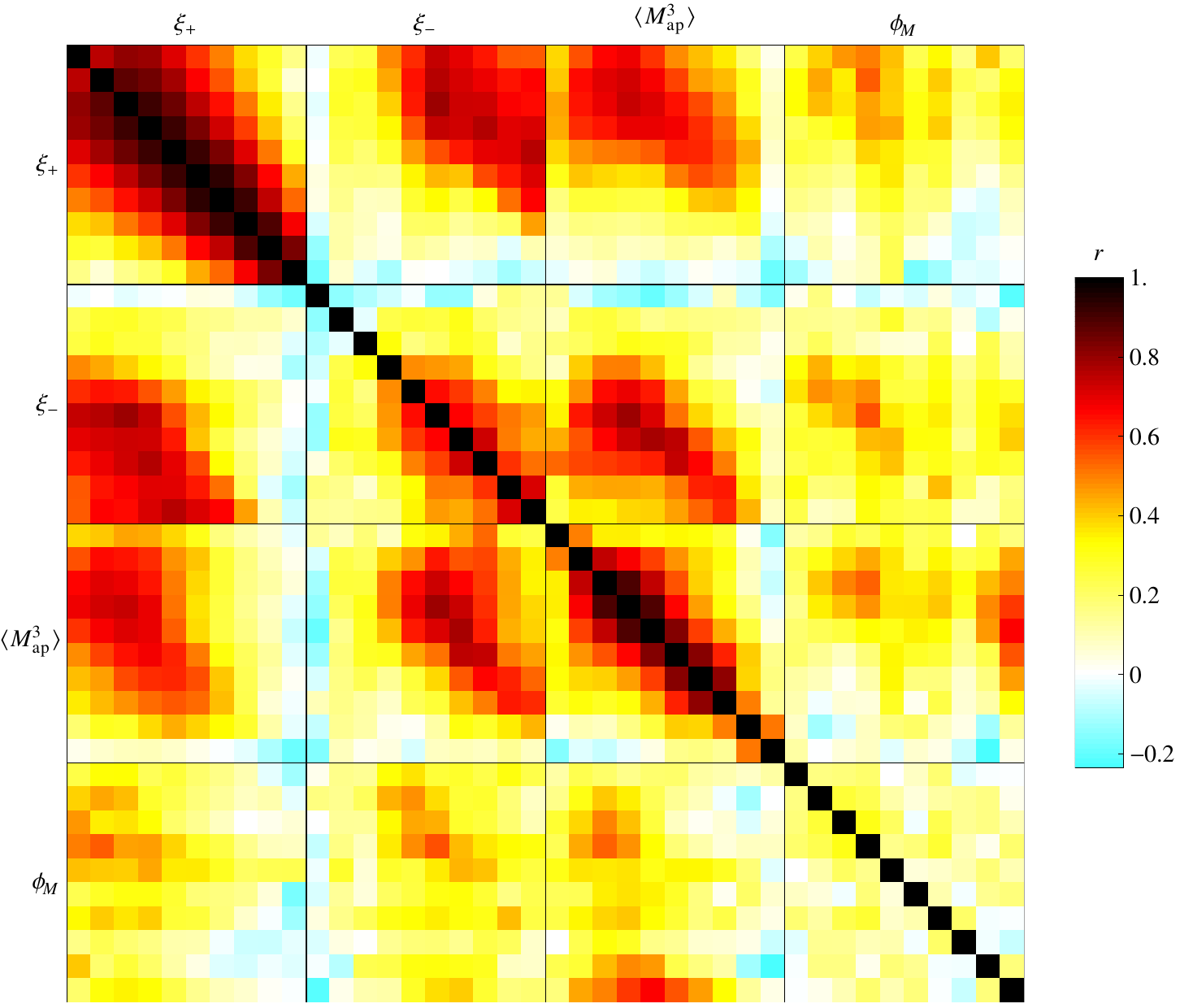}}
\caption{
\label{fig:correlation_matrix}
Correlation matrix quantifying the correlation between the entries of the full data vector, containing the shear correlations functions $\xi_\pm(\theta)$ measured in 10 bins of pair separation $\theta$, the aperture statistics $\bEV{\Map^3}(\vartheta)$ measured for 10 filter scales $\theta$, and the shear peak abundance $\dpa(M)$ measured in 10 bins of peak mass $M$. Within each class of observables, filter masses, scales or separations increase from left to right and from top to bottom. Darker colors indicate larger absolute correlation.
}
\end{figure*}

The error analysis in the previous section was restricted to the autovariance of the lensing observables $\xi_\pm(\theta)$, $\bEV{\Map^3}(\vartheta)$, and $\dpa(M)$. Our lensing simulations allow us to estimate the full covariance between these observables, quantified by the covariance matrix $\matrb{C}_{\text{d}}$. For an analysis of the correlation between the observables, we consider the correlation matrix
\begin{equation}
	\matrb{R}_\text{d} = \left( \frac{\matrb{C}_{\text{d},ij}}{\sqrt{\matrb{C}_{\text{d},ii}\matrb{C}_{\text{d},jj}}}\right)_{i,j=1}^{N_\text{d}}.
\end{equation}

The correlation matrix is shown in Fig.~\ref{fig:correlation_matrix}. All $\xi_+$-bins with similar separations $\vartheta$ are strongly correlated with each other. The correlation among the $\xi_-$-bins and among the $\bEV{\Map^3}$-bins is slightly smaller. There is also a substantial correlation between $\xi_+$, $\xi_-$, and $\bEV{\Map^3}$. Such large correlations are expected for surveys like the one considered here, where the cosmic variance is the dominant source of uncertainties.

In contrast, there is little correlation between different bins of $\dpa$. The correlation between $\dpa$ and the other lensing observables is not very large either. This indicates that by combining the shear peak statistics with the other probes, one can obtain substantially more information than one could get from either shear peaks or, e.g., cosmic shear correlations alone \citep[as advocated by][]{DietrichHartlap2010}.

\subsection{Derivatives of the lensing observables}
\label{sec:results:derivatives}

\begin{figure*}
\centerline{\includegraphics[width=1.0\linewidth]{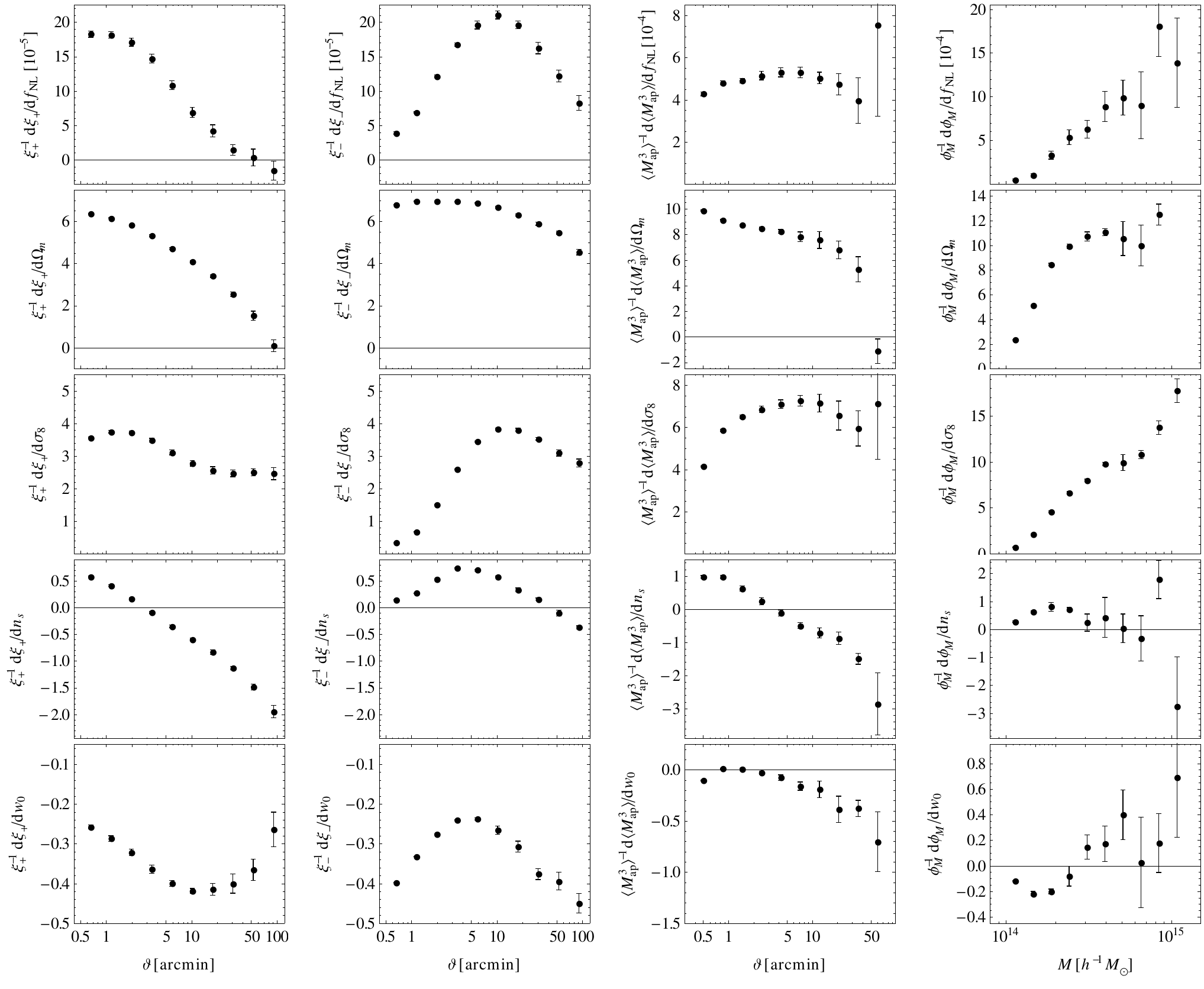}}
\caption{
\label{fig:relative_derivatives}
Derivatives of the cosmic shear correlations $\xi_\pm$, the aperture statistics $\bEV{M_\text{ap}^3}$, and the shear peak abundance $\dpa$ with respect to the non-Gaussianity parameter $\fNL$, the mean matter density $\Omega_\text{m}$, the matter power spectrum normalization $\sigma_8$ and the spectral index $n_\text{s}$, and the equation-of-state parameter $w_0$ of the dark energy, as estimated from our simulations (points with error bars indicating uncertainties on the estimate).
}
\end{figure*}

\begin{figure*}
\centerline{\includegraphics[width=1.0\linewidth]{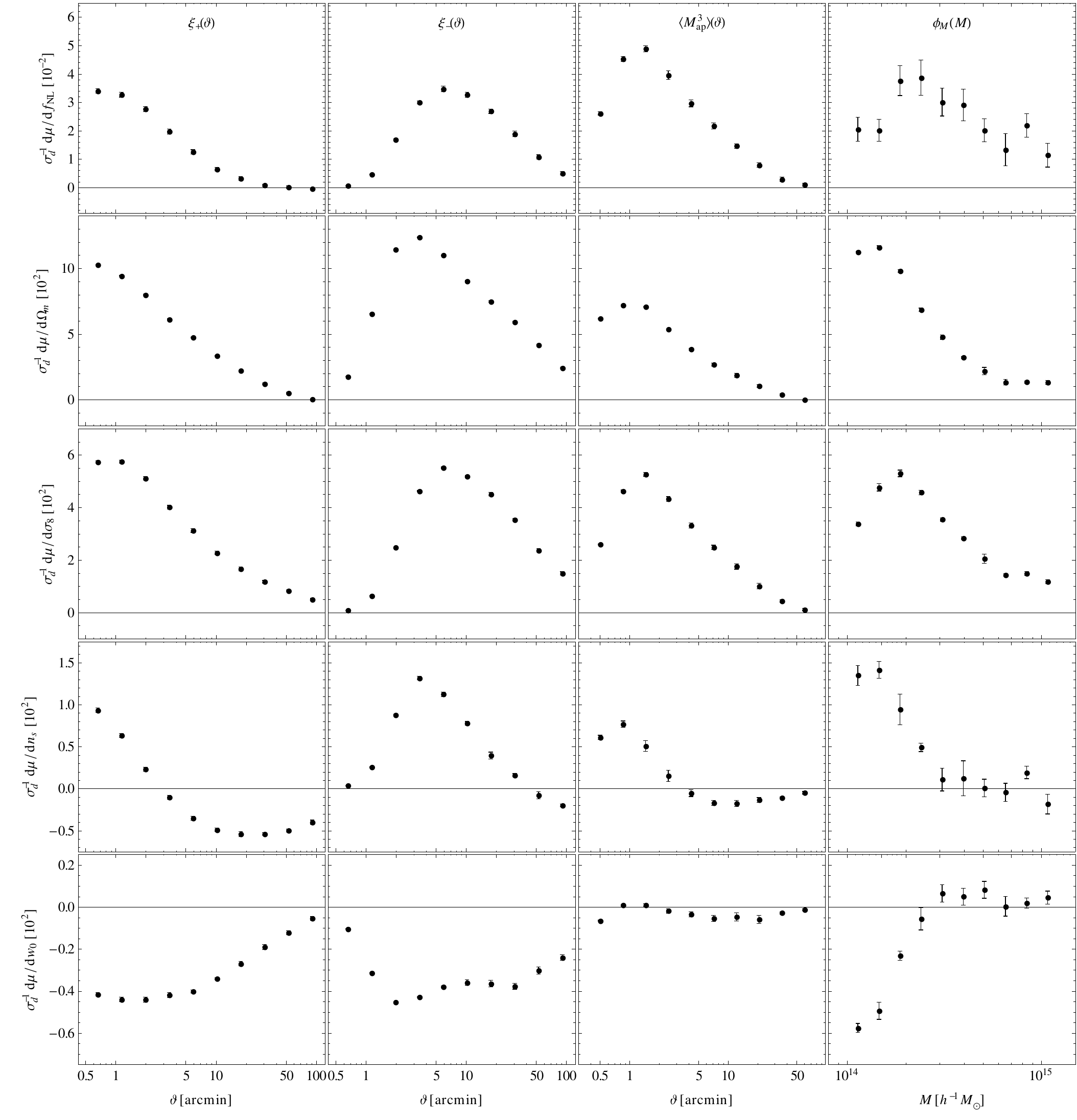}}
\caption{
\label{fig:relative_derivatives_std_dev}
Derivatives of the cosmic shear correlations $\xi_\pm$ (1st and 2nd column), the aperture statistics $\bEV{M_\text{ap}^3}$ (3rd column), and the shear peak abundance $\dpa$ (4th column) with respect to the non-Gaussianity parameter $\fNL$ (top row), the mean matter density $\Omega_\text{m}$ (2nd row), the matter power spectrum normalization $\sigma_8$ (3rd row) and the spectral index $n_\text{s}$ (4th row), and the equation-of-state parameter $w_0$ of the dark energy (5th row) relative to the statistical errors $\sigma_{\text{d}}$ on the observables predicted for a large future survey (points with error bars indicating uncertainties stemming from our noisy estimates of the derivatives).
}
\end{figure*}

In order to translate the measurement uncertainties and correlations in the lensing observables into uncertainties and correlations in the cosmological parameters, we need to know how the observables respond to changes in the parameters. As apparent from Eq.~\eqref{eq:simple_likelihood_parameter_covariance}, the derivatives of the model prediction with respect to the parameters are sufficient for this.

The derivatives we estimate from our simulations via finite difference are shown in Fig.~\ref{fig:relative_derivatives}. The Figure also shows the statistical uncertainties on our results for the derivatives stemming from our limited number of realizations. In most cases, the derivatives bear a relatively small statistical uncertainty. Only for $\bEV{M_\text{ap}^3}$ and $\dpa$, the uncertainties in the derivatives reach a magnitude similar to the derivatives themselves on larger aperture and mass scales.

In order to understand how much each observable contributes to the parameter constraints, and how much the uncertainties in the derivatives might affect the estimated parameter errors, we consider the individual contributions to the sum in Eq.~\eqref{eq:simple_likelihood_parameter_errors}. These contributions, the derivatives rescaled by the expected statistical errors on the observables due to shape noise and cosmic variance, can be considered as measures of differential signal-to-noise. They are shown in Fig.~\ref{fig:relative_derivatives_std_dev}.

The scales with the largest contributions to constraints on $\fNL$ from $\xi_+$ and $\xi_-$ are \mbox{$\sim1\,\arcmint$} and \mbox{$\sim10\,\arcmint$}, which are the scales where the relative changes in $\xi_+$ and $\xi_-$ with changing $\fNL$ are largest. For $\bEV{\Map^3}$, the largest contributions are from scales $\sim1\,\arcmint$, where the relative errors are smallest. For $\dpa$, the largest contributions to constraints on $\fNL$ are not from large peak masses, where the relative change with changing $\fNL$ is largest, but from intermediate peak masses due to the much smaller noise.\footnote{
Note however that this simple consideration ignores correlations between bins.
}

The results for the contributions to the constraints on the other cosmological parameters are similar. In general, most constraints come from small and intermediate scales, with the largest angular and mass scales contributing relatively little. In particular, the peak counts in the small mass range appear to provide the largest contributions to constraints on the DE EOS parameter $w_0$.

\subsection{Cosmological constraints}
\label{sec:results:constraints}

In this section, we discuss the cosmological constraints expected from measuring and analyzing the cosmic shear correlation functions $\xi_\pm(\vartheta)$, the third-order aperture statistics $\bEV{\Map^3}(\vartheta)$, and the shear peak statistics $\dpa(M)$ in a large future WL survey.

We restrict the analysis to flat CDM cosmologies with a dark energy with a time-independent equation of state and local-type primordial non-Gaussianity, which is the class of models covered by our suite of simulations. The considered parameters are the non-Gaussianity parameter $\fNL$, the mean matter density $\Omega_\text{m}$, the matter power spectrum normalization $\sigma_8$ and spectral index $n_\text{s}$, and the dark-energy equation-of-state parameter $w_0$. Here, we concentrate on the conditional and marginal parameter constraints, as well as possible biases due to ignoring non-vanishing non-Gaussianity.
The full posterior parameter covariances for the considered probes are listed in Appendix~\ref{sec:appendix:parameter_covariance_matrices}.

\subsubsection{Conditional constraints on cosmological parameters}
\label{sec:results:constraints:conditional_constraints}

\begin{table}
\center
\caption{
\label{tab:conditional_constraints}
Conditional constraints on cosmological parameters from a large future weak lensing survey: estimated conditional posterior standard deviations $\sigma_\text{c}$ of the non-Gaussianity parameter $\fNL$, the mean matter density $\Omega_\text{m}$, the normalization $\sigma_8$ and the spectral index $n_\text{s}$ of the matter power spectrum, and the equation-of-state parameter $w_0$ of the dark energy, obtained when using the cosmic shear correlations $\xi_\pm$, the aperture statistics $\bEV{M_\text{ap}^3}$, or the shear peak abundance $\dpa$ either alone or combined with each other.
}
\begin{tabular}{l r r r r r r}
$\sigma_{\text{c}}$ from\textbackslash of & $\fNL$ & $\Omega_\text{m}$ & $\sigma_8$ & $n_\text{s}$  &  $w_0$ \\
\hline
$\xi_\pm$                                 & 21     & 0.0006            & 0.0014     & 0.0046        & 0.0149 \\
$\bEV{M_\text{ap}^3}$                     & 20     & 0.0012            & 0.0019     & 0.0089        & 0.0920 \\
$\dpa$                                    & 17     & 0.0006            & 0.0013     & 0.0053        & 0.0136 \\
\hline
$\xi_\pm$ + $\bEV{M_\text{ap}^3}$         & 17     & 0.0005            & 0.0013     & 0.0037        & 0.0103 \\
$\xi_\pm$ + $\dpa$                        & 16     & 0.0006            & 0.0013     & 0.0043        & 0.0104 \\
$\bEV{M_\text{ap}^3}$ + $\dpa$            & 17     & 0.0006            & 0.0013     & 0.0051        & 0.0128 \\
\hline
all WL                                    & 14     & 0.0005            & 0.0012     & 0.0038        & 0.0089 \\ 
\end{tabular}
\end{table}

\begin{figure*}
\centerline{\includegraphics[width=0.8\linewidth]{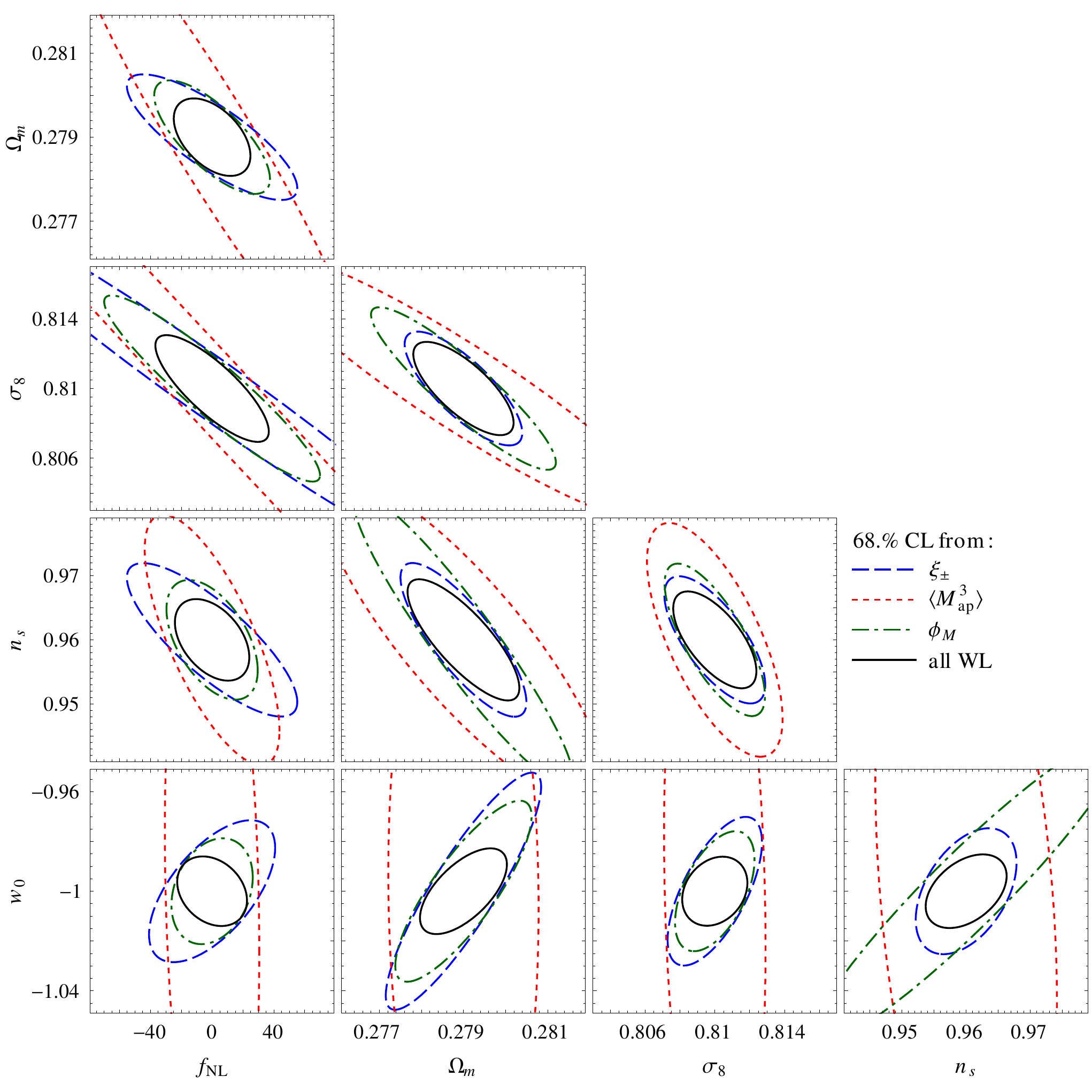}}
\caption{
\label{fig:conditional_parameter_error_contours}
Joint conditional constraints on pairs of cosmological parameters from a large future weak lensing survey. Considered parameters are the non-Gaussianity parameter $\fNL$, the mean matter density $\Omega_\text{m}$, the normalization $\sigma_8$ and the spectral index $n_\text{s}$ of the matter power spectrum, and the equation-of-state parameter $w_0$ of the dark energy. Shown are the estimated 68\% confidence contours obtained from the cosmic shear correlations $\xi_\pm$ (dashed lines), the aperture statistics $\bEV{M_\text{ap}^3}$ (dotted lines), the shear peak abundance $\dpa$ (dash-dotted lines), all of these WL probes together (solid lines).
}
\end{figure*}

The conditional posterior uncertainties expected from our model survey on individual cosmological parameters, when flat priors are assumed, are listed in Table~\ref{tab:conditional_constraints}. When all other cosmological parameters are known, the cosmic shear correlations $\xi_\pm$, the third-order aperture statistics $\bEV{\Map^3}$, and the shear peak abundance $\dpa$ each yield 1-sigma constraints of $\Delta\fNL \sim 20$, with the smallest error $\Delta\fNL \approx 17$ from $\dpa$. When all WL probes are combined, the conditional error $\Delta\fNL \approx 14$.

The cosmic shear correlation functions and shear peak abundance also provide similar conditional constraints on the other cosmological parameters $\Omega_\text{m}$, $\sigma_8$, $n_\text{s}$, and $w_0$. The parameter constraints from the aperture statistics are generally weaker than those from $\xi_\pm$ or $\dpa$.
By combining two or all three of the considered WL probes, the errors on $\Omega_\text{m}$, $\sigma_8$, $n_\text{s}$, and $w_0$ can also be improved. The relative conditional constraints on these parameters expected from a combined analysis of all the WL probes are $\sim 0.1 \% - 1\%$, as shown in Table~\ref{tab:conditional_constraints}.

The joint conditional constraints for pairs of cosmological parameters are illustrated in Fig.~\ref{fig:conditional_parameter_error_contours}. For all considered WL probes, $\fNL$ shows strong \mbox{(anti-)}correlations with the other considered cosmological parameters. (Anti-)correlations are also visible between the other parameters, e.g. between $\sigma_8$ and $\Omega_\text{m}$. The constraints from combining the WL probes benefit substantially from the complementary constraints provided by the different WL probes. Combining the information from the different probes also reduces the parameter correlations.

\subsubsection{Marginalized constraints on cosmological parameters}
\label{sec:results:constraints:marginal_constraints}

\begin{table}
\center
\caption{
\label{tab:marginal_constraints}
Marginal constraints on cosmological parameters from a large future weak lensing survey: estimated posterior standard deviations $\sigma_{\text{po}}$ of the non-Gaussianity parameter $\fNL$, the mean matter density $\Omega_\text{m}$, the normalization $\sigma_8$ and the spectral index $n_\text{s}$ of the matter power spectrum, and the equation-of-state parameter $w_0$ of the dark energy, obtained when using the cosmic shear correlations $\xi_\pm$, the aperture statistics $\bEV{M_\text{ap}^3}$, or the shear peak abundance $\dpa$ either alone, all together, or all together in conjunction with results from a CMB experiment like \satellitename{Planck}.
}
\begin{tabular}{l r r r r r r}
$\sigma_{\text{po}}$ from\textbackslash of & $\fNL$ & $\Omega_\text{m}$ & $\sigma_8$ & $n_\text{s}$  &  $w_0$ \\
\hline
$\xi_\pm$                                  & 213    & 0.0063            & 0.0118     & 0.0258        & 0.1119 \\
$\bEV{M_\text{ap}^3}$                      & 444    & 0.0102            & 0.0423     & 0.0456        & 0.1873 \\
$\dpa$                                     &  69    & 0.0084            & 0.0147     & 0.0316        & 0.1170 \\ 
\hline
$\xi_\pm$ + $\bEV{M_\text{ap}^3}$          & 141    & 0.0042            & 0.0083     & 0.0127        & 0.0770 \\
$\xi_\pm$ + $\dpa$                         &  60    & 0.0020            & 0.0039     & 0.0084        & 0.0278 \\
$\bEV{M_\text{ap}^3}$ + $\dpa$             &  63    & 0.0040            & 0.0072     & 0.0157        & 0.0385 \\
\hline
all WL                                     &  53    & 0.0017            & 0.0035     & 0.0067        & 0.0263 \\
all WL + CMB                               &  51    & 0.0014            & 0.0034     & 0.0036        & 0.0252 \\
\end{tabular}
\end{table}

\begin{figure*}
\centerline{\includegraphics[width=0.8\linewidth]{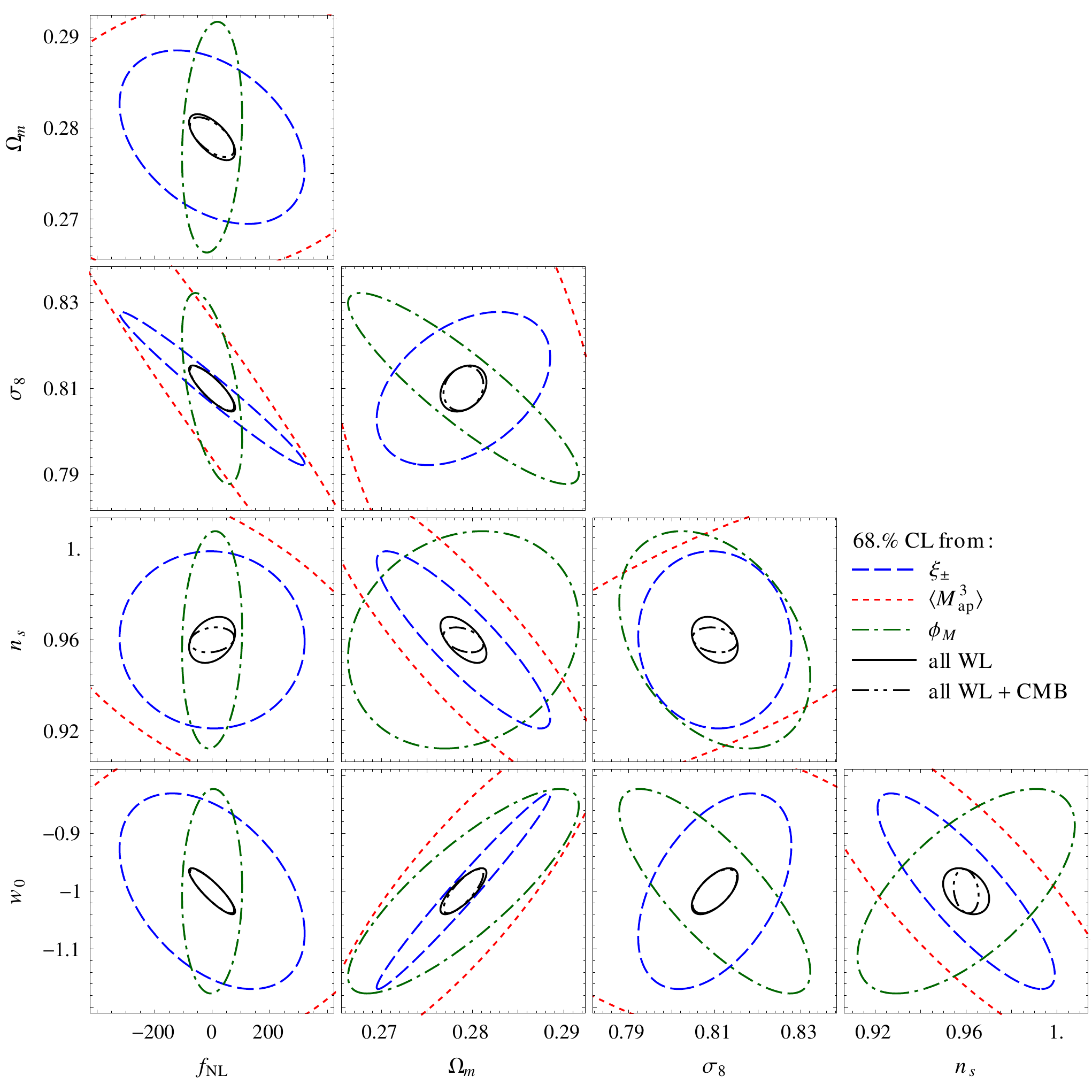}}
\caption{
\label{fig:marginal_parameter_error_contours}
Joint marginal constraints on pairs of cosmological parameters from a large future weak lensing survey. Considered parameters are the non-Gaussianity parameter $\fNL$, the mean matter density $\Omega_\text{m}$, the normalization $\sigma_8$ and the spectral index $n_\text{s}$ of the matter power spectrum, and the equation-of-state parameter $w_0$ of the dark energy. Shown are the estimated 68\% confidence contours obtained from the cosmic shear correlations $\xi_\pm$ (dashed lines), the aperture statistics $\bEV{M_\text{ap}^3}$ (dotted lines), the shear peak abundance $\dpa$ (dash-dotted lines), all of these WL probes together (solid lines), or all WL probes combined with results from a \satellitename{Planck}-like CMB experiment.
}
\end{figure*}

The marginal posterior standard deviations on the cosmological parameters, assuming flat priors, are shown in Table~\ref{tab:marginal_constraints}. The marginal constraints on $\fNL$ from $\bEV{M_\text{ap}^3}$ alone are so large  ($\Delta\fNL \approx 500$) that they are practically useless due to the relatively large uncertainties in the other cosmological parameters and their correlation with $\fNL$. The marginal errors on $\fNL$ from $\xi_\pm$ ($\Delta\fNL \approx 200$) or $\dpa$ ($\Delta\fNL \approx 70$) alone are much smaller than the constraints from $\bEV{M_\text{ap}^3}$ thanks to the thighter constraints on the other cosmological parameters. All WL probes combined yield a marginal error $\Delta\fNL \approx 50$, which is somewhat larger than current constraints from CMB bispectra ($\Delta\fNL \approx 20$).

The marginal errors from either the shear correlations, the aperture statistics, or the peak counts on the other cosmological parameters are $\sim1\% - 20\%$ of the fiducial values. However, combining the information from cosmic shear and the shear peak abundance yields much more competitive marginal constraints, with uncertainties much smaller than those from any of the individual WL probes. This is in accordance with findings by \citet{DietrichHartlap2010}, whose study however was limited to $\Omega_{\text m}$ and $\sigma_8$. An additional improvement can be obtained by also adding information from the aperture statistics. The great improvement reached by combining the three WL probes compared to only analyzing any single WL probe is also evident in the joint marginal constraints shown Fig.~\ref{fig:marginal_parameter_error_contours}.

Table~\ref{tab:marginal_constraints} also lists constraints from combining the results from the WL probes with priors that CMB temperature and E-mode polarization correlation measurements with an experiment like \satellitename{Planck} could provide for the cosmological parameters other than $\fNL$ (see Appendix~\ref{sec:appendix:planck_fisher_matrix} for a computation of the prior Fisher matrix). However, the improvement on the marginal uncertainties on $\fNL$ predicted from this combination is very small. The reduction expected for the uncertainties in the other parameters is not large either except for the spectral index $n_\text{s}$.

To assess the robustness of our estimated parameter constraints to changes of the realizations entering the covariance, we also compute the cosmological parameter constraints using the data covariance estimated from the 128 simulated fields of the fiducial \softwarename{zHORIZON} runs. The resulting constraints are larger or smaller by up to few tens of percent, with a slight trend toward larger uncertainties for the results based on the \softwarename{zHORIZON} covariances. For example, the marginal constraints from combining all three WL probes become: $\Delta \fNL = 50$, $\Delta \Omega_\text{m} = 0.0018$, $\Delta \sigma_8 = 0.0038$, $\Delta n_\text{s} = 0.0084$, $\Delta w_0 = 0.028$. In comparison with the results listed in Table \ref{tab:marginal_constraints}, the largest difference (a 25\% increase) is observed for $n_\text{s}$.

\subsection{Biases induced by neglecting non-Gaussianity}
\label{sec:results:biases}

\begin{table}
\center
\caption{
\label{tab:biases}
Biases in cosmological parameters induced by assuming the `wrong' value for the primordial non-Gaussianity $\fNL$ in the analysis of a large future weak lensing survey: differences $\Delta\Omega_\text{m}$, $\Delta\sigma_8$, $\Delta n_\text{s}$, and $\Delta w_0$ between the expected posterior means and the `true' values of the mean matter density $\Omega_\text{m}$, the normalization $\sigma_8$, the spectral index $n_\text{s}$, and the equation-of-state parameter $w_0$, resp., relative to the difference $\Delta\fNL$ between the `true' value of the primordial non-Gaussianity $\fNL$ and its value assumed in the analysis.
}
\begin{tabular}{l r r r r}
& $\dfrac{\Delta\Omega_\text{m}}{\Delta\fNL}$ & $\dfrac{\Delta\sigma_8}{\Delta\fNL}$ & $\dfrac{\Delta n_\text{s}}{\Delta\fNL}$ & $\dfrac{\Delta w_0}{\Delta\fNL}$ \\
\hline
$\xi_\pm$                          & $-1\times10^{-5}$   & $-5\times10^{-5}$   & $-2\times10^{-6}$   & $-2\times10^{-4}$ \\
$\bEV{M_\text{ap}^3}$              & $ 3\times10^{-6}$   & $-9\times10^{-5}$   & $-6\times10^{-5}$   & $ 9\times10^{-5}$ \\
$\dpa$                             & $ 2\times10^{-5}$   & $-1\times10^{-4}$   & $ 5\times10^{-5}$   & $ 7\times10^{-5}$ \\
\hline
$\xi_\pm$ + $\bEV{M_\text{ap}^3}$  & $-3\times10^{-5}$   & $-6\times10^{-5}$   & $ 7\times10^{-5}$   & $-5\times10^{-4}$ \\
$\xi_\pm$ + $\dpa$                 & $-2\times10^{-5}$   & $-5\times10^{-5}$   & $ 6\times10^{-5}$   & $ 4\times10^{-4}$ \\
$\bEV{M_\text{ap}^3}$ + $\dpa$     & $ 5\times10^{-6}$   & $-8\times10^{-5}$   & $-7\times10^{-5}$   & $-3\times10^{-4}$ \\
\hline
all WL                             & $-2\times10^{-5}$   & $-6\times10^{-5}$   & $ 4\times10^{-5}$   & $-4\times10^{-4}$ \\
\end{tabular}
\end{table}

\begin{figure*}
\centerline{\includegraphics[width=0.62\linewidth]{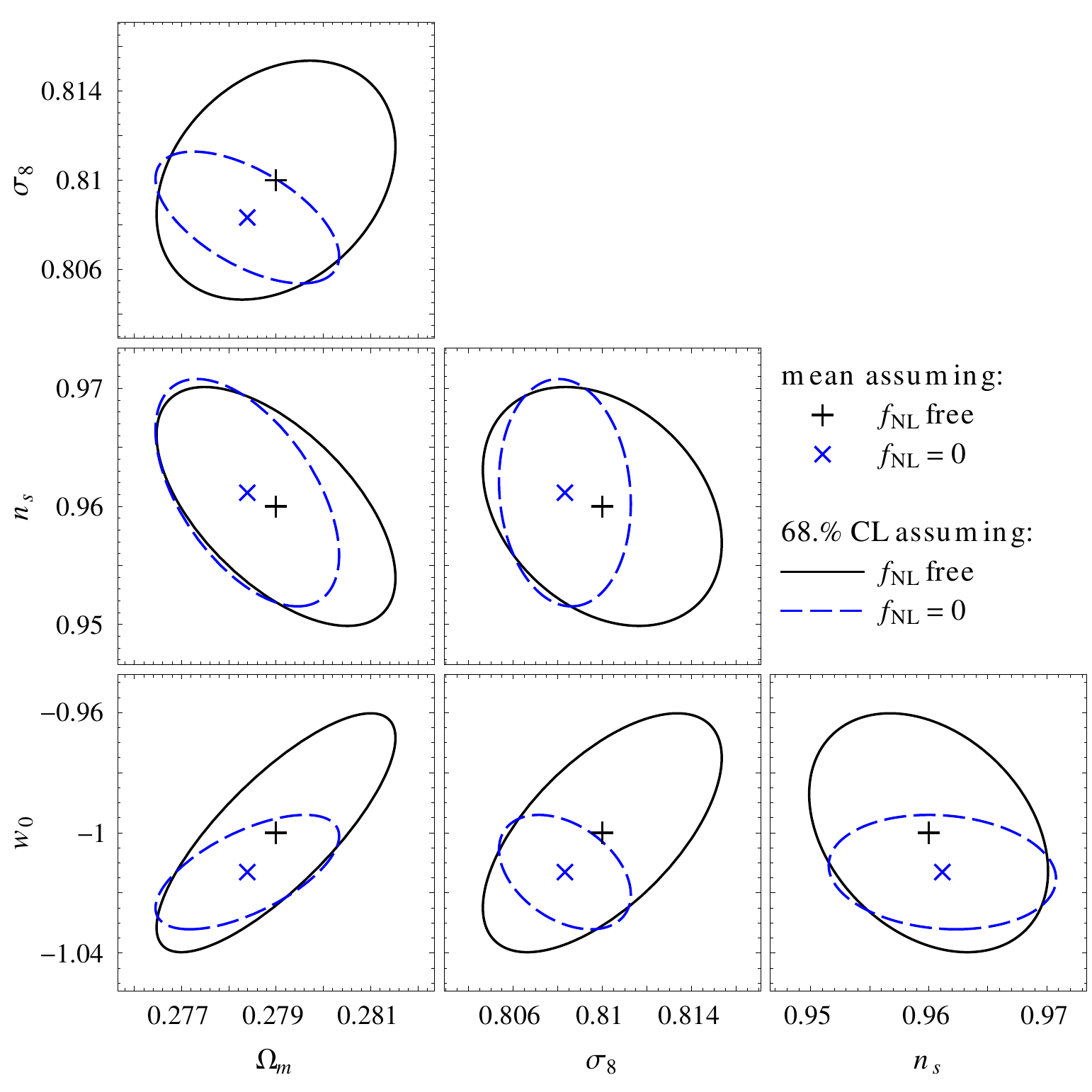}}
\caption{
\label{fig:biased_parameter_error_contours}
Joint marginal constraints on pairs of cosmological parameters from a large future weak lensing survey: posterior means (symbols) and 68\% confidence contours (lines) for the mean matter density $\Omega_\text{m}$, the normalization $\sigma_8$ and the spectral index $n_\text{s}$, and the equation-of-state parameter $w_0$ obtained for a cosmology with `true' non-Gaussianity parameter $\fNL = 30$ when either allowing for $\fNL$ to vary (+, solid lines), or assuming $\fNL = 0$ ($\times$, dashed lines) in the joint WL analysis.
}
\end{figure*}

The constraints on the cosmological parameters predicted for a future all-sky weak lensing survey imply substantial correlations between $\fNL$ and other cosmological parameters. This raises the question whether neglecting a possible non-Gaussianity at the level of current constraints on $\fNL$ might induce a serious bias in the other cosmological parameters. Such a bias would ruin the determination of the cosmological parameters if the shift in the inferred posterior parameter confidence regions induced by neglecting non-Gaussianity were comparable to or larger than the inferred parameter uncertainties.

We study this possible bias on the basis of Eqs.~\eqref{eq:conditional_for_Gaussian_posterior} to \eqref{eq:conditional_covariance_for_Gaussian_posterior}, assuming flat priors for the cosmological parameters and the likelihood computed from our simulations. Using Eq.~\eqref{eq:conditional_mean_for_Gaussian_posterior}, we can compute how the inferred posterior means of the other cosmological parameters respond to a `wrong' choice (i.e. assuming a value that differs from the `true' value) for $\fNL$. This response is listed in Table~\ref{tab:biases} as ratio between the induced systematic error in the cosmological parameters and the systematic error $\Delta\fNL$ in the non-Gaussianity parameter.

For illustrative purposes, we consider a scenario where the `true' non-Gaussianity $\fNL = 30$ \citep[which is roughly the current best guess,][]{KomatsuEtal2011short}. The joint marginal posterior constraints on the parameters $\Omega_\text{m}$, $\sigma_8$, $n_\text{s}$, and $w_0$ resulting from a joint analysis of $\xi_\pm$, $\bEV{M_\text{ap}^3}$, and $\dpa$ are shown in Fig.~\ref{fig:biased_parameter_error_contours} for two cases. In the first case, $\fNL$ is considered a parameter to be constrained from the data along with the other parameters. For our simple model, the resulting posterior means coincide with the `true' values of the parameters. The second case assumes $\fNL=0$ in the analysis. This results in different means and covariances inferred for the parameters. The resulting biases are noticeable compared to the parameter uncertainties estimated for the full model including $\fNL$. Moreover, the analysis ignoring a possible non-Gaussianity yields not only biased estimates, but also far too optimistic uncertainties for the remaining parameters, in particular for $\sigma_8$ and $w_0$. Thus, the parameter biases are comparable to the inferred uncertainties in that case. As a result, the confidence regions inferred by an analysis ignoring non-Gaussianity do not contain the `true' values of the cosmological parameters.

\section{Summary and discussion}
\label{sec:summary}

Measurements of the level of non-Gaussianity in the primordial matter density fluctuations provide valuable information for constraining models of inflation and the origin of cosmic structure. Currently the best constraints on primordial non-Gaussianity are from CMB experiments probing matter fluctuations at very high redshifts. However, lower-redshift observables could also provide competitive constraints.

In this paper, we investigated how well a large weak lensing survey probing the matter distribution at redshifts $0 < z \lesssim 1$, can constrain the level $\fNL$ of primordial non-Gaussianity of the local type. We employed a large set of $N$-body simulations of structure formation in flat $w$CDM cosmologies, through which we performed full gravitational ray-tracing to simulate weak lensing observations similar to those proposed for \satellitename{Euclid} and LSST. With these WL simulations, we studied the influence of the cosmological parameters $\fNL$, $\Omega_\text{m}$, $\sigma_8$, $n_\text{s}$, and $w_0$ on the cosmic shear correlation functions $\xi_+$ and $\xi_-$, the third-order aperture mass statistics $\bEV{\Map^3}$, and the shear peak abundance $\dpa$. Furthermore, we quantified the uncertainties and covariances of these WL probes, and the resulting constraints on the cosmological parameters.

We found that if all other cosmological parameters were known accurately, each of the considered WL probes could provide constraints $\Delta\fNL \sim 20$ on the non-Gaussianity parameter $\fNL$. The constraints were much weaker ($\Delta\fNL \sim 100 - 500$), if the other cosmological parameters had to be constrained using the same WL data. However, the uncertainties in $\fNL$ could be reduced by combining information from the different WL probes.

A joint analysis of the WL probes also yields much smaller uncertainties for the other cosmological parameters than may be obtained from any of the individual WL probes. When combining the probes, the expected marginal uncertainties at 68\% confidence level on the cosmological parameters were: $\Delta \fNL \sim 50$, $\Delta \Omega_\text{m} \sim 0.002$, $\Delta \sigma_8 \sim 0.004$, $\Delta n_\text{s} \sim 0.007$, and $\Delta w_0 \sim 0.03$. 

At this level of accuracy, neglecting non-vanishing non-Gaussianity could induce a noticeable bias in the determination of other cosmological parameters, in particular for the matter power spectrum normalization $\sigma_8$ or the DE EOS parameter $w_0$. Moreover, neglecting a possible non-Gaussianity in the analysis may lead to far too optimistic estimates for the parameter errors.

Our results for the expected conditional errors on $\fNL$ from cosmic shear are in agreement with the findings of \citet{FedeliMoscardini2010}, who estimated errors of a few tens considering the weak lensing power spectra. Our results for the expected constraints from shear peaks are somewhat less optimistic than the errors predicted by \citet{MaturiFedeliMoscardini2011}. This might be attributed to our more realistic predictions for the peak function, which we measured directly from simulated WL maps.

The analysis in Section~\ref{sec:results:derivatives} showed that a large part of the parameter constraints stems from the lensing signals measured on small and intermediate angular or mass scales. This highlights the importance of improving the understanding of structure formation on these smaller scales, which is required for accurate model predictions. Moreover, an improved understanding of the shape noise is needed for accurate predictions of the shear peak statistics at these scales.

Our error analysis in Section~\ref{sec:results:error_analysis} showed that for large future WL surveys, the statistical errors on the cosmic shear correlations and the aperture statistics are cosmic-variance dominated. This implies that for these probes, reducing the shape noise, e.g. by increasing the survey depth, will not reduce the statistical errors much.

An obvious extension of this work is to take tomographic information into account. Tomography might not significantly improve the errors on $\fNL$ arising from two-point statistics like the convergence power spectrum \citep[see][]{FedeliMoscardini2010}, but it could improve the constraints from the peak abundance. It will also significantly reduce the errors on other parameters, such as the equation-of-state parameters of the Dark Energy. Such a study probably requires more simulations to obtain reliable covariances for the large data vectors arising in WL tomography.

One should also investigate more the constraints one may obtain from combining the WL probes with other probes of large scale structure \citep[see][]{GiannantonioEtal2012}. This will require simulations that include predictions for optical properties of galaxies or X-ray properties of clusters.
Finally, a comprehensive future analysis should also consider other types of primordial non-Gaussianity and other cosmological parameters impacting structure formation.

\section*{Acknowledgments}
We thank Alan Heavens, Carlos Cunha, and Peter Schneider for helpful discussions. We also thank Albert Stebbins for the constructive criticsm. SH and LM acknowledge support by the Deutsche Forschungsgemeinschaft (DFG) through the Priority Programme 1177 `Galaxy Evolution' (SCHN 342/6 and WH 6/3), through the Transregional Collaborative Research Centre TRR 33 `The Dark Universe'. and through the grant MA 4967/1-1. SH also acknowledges support by the National Science Foundation (NSF) grant number AST-0807458-002. RES was partly supported by the Swiss National Foundation under contract 200021-116696/1 and the WCU grant R32-2008-000-10130-0. RES also acknowledges support from a Marie Curie Reintegration Grant and the Alexander von Humboldt Foundation. VD acknowledges support by the Swiss National Foundation and by the University of Z{\"u}rich through FK UZH 57184001.




\appendix

\section{Planck Fisher matrix}
\label{sec:appendix:planck_fisher_matrix}

\subsection{Computing the CMB matrix}

In computing the Planck Fisher matrix we follow the methodology
described in \citet{EisensteinHuTegmark1999} and for the specific
implementation we follow \citet{TakadaJain2009}. We thus assume that
the CMB temperature and polarization spectra constrain 9 parameters,
and for our calculations we set their fiducial values to be those from
the recent WMAP7 analysis \citep{KomatsuEtal2011short}. The fiducial
parameters are: dark energy EOS parameters $w_0=-1.0$ and $w_a=0.0$;
the density parameter for dark energy $\Omega_\mathrm{DE}=0.7274$; the
CDM and baryon density parameters scaled by the square of the
dimensionless Hubble parameter $w_\mathrm{c}=\Omega_\mathrm{c} h^2=0.1125$
and $w_\mathrm{b}=\Omega_\mathrm{b}h^2=0.0226$
($h=H_0/[100 \mathrm{kms^{-1}Mpc^{-1}}]$); and the primordial spectral
index of scalar perturbations $n_\mathrm{s}=0.963$; the primordial amplitude of
scalar perturbations $A_\mathrm{s}=2.173\times10^{-9}$; the running of the
spectral index $\alpha=0.0$; and the optical depth to the last
scattering surface $\tau=0.087$. Hence we may write our vector of
parameters:
\begin{equation}
\transposed{\vect{p}_-}  =  \transposed{(w_0,w_a,\Omega_\mathrm{DE},w_\mathrm{c},w_\mathrm{b},\tau,n_\mathrm{s},A_\mathrm{s},\alpha)}.
\end{equation}
Whilst the CMB does constrain $\fNL$ through the temperature
and polarization bispectra, here we are only interested in assessing
the potential of weak lensing to place constraints on $\fNL$.
We therefore extend the vector $\vect{p}_-$ to the 10-dimensional vector:
\begin{equation}
\vect{p}_- \mapsto  \vect{p} = \transposed{(w_0,\ldots,\alpha,\fNL)}
.
\end{equation}

The CMB Fisher matrix can be written as \citep{EisensteinHuTegmark1999}:
\def\Fisher{{\mathbf{F}}}
\begin{equation} \Fisher_{p_\alpha p_\beta}=
\sum_\ell \sum_{X,Y} 
\frac{\partial C_{\ell,X}}{\partial p_\alpha}
\mathrm{Cov}^{-1}\left[C_{\ell,X},C_{\ell,Y}\right]
\frac{\partial C_{\ell,Y}}{\partial p_\beta}\ ,
\end{equation}
where $\{X,Y\}\in \{\mathrm{TT},\,\mathrm{EE},\,\mathrm{TE}\}$, where $C_{\ell,\rm
  TT}$ is the temperature power spectrum, $C_{\ell,\rm EE}$ is the E-mode
polarization power spectrum, $C_{\ell,\rm TE}$ is the temperature-E-mode
polarization cross-power spectrum. We have been conservative and
assumed that there will be no significant information from the
$C_{\ell,\rm BB}$, the B-mode polarization power spectrum. We compute all
CMB spectra using {\tt CAMB} \citep{LewisChallinorLasenby2000} and use the
additional module for time evolving dark energy models
\citep{HuSawicki2007}. We include information from all multipoles in
the range ($2< \ell <1500$).

The covariance matrices for these observables are:
\begin{align} 
\cov\left[C_{\ell,\TT},C_{\ell,\TT}\right] & = \frac{1}{f_\mathrm{sky}}
\frac{2}{2\ell+1}\left[C_{\ell,\TT}+N_{\ell,\TT}\right]^2 \ ;\\ 
\cov\left[C_{\ell,\TT},C_{\ell,\EE}\right] & = \frac{1}{f_\mathrm{sky}}
\frac{2}{2\ell+1}  C_{\ell,\TE}^2 \ ;                \\
\cov\left[C_{\ell,\TT},C_{\ell,\TE}\right] & = \frac{1}{f_\mathrm{sky}}
\frac{2}{2\ell+1} C_{\ell,\TE}\left[C_{\ell,\TT}+N_{\ell,\TT}\right]\ ;\\
\cov\left[C_{\ell,\EE},C_{\ell,\EE}\right] & = \frac{1}{f_\mathrm{sky}}
\frac{2}{2\ell+1} \left[C_{\ell,\EE}+N_{\ell,\EE}\right]^2\ ;\\
\cov\left[C_{\ell,\EE},C_{\ell,\TE}\right] & = \frac{1}{f_\mathrm{sky}}
\frac{2}{2\ell+1}C_{\ell,\TE}\left[C_{\ell,\EE}+N_{\ell,\EE}\right] \ ;\\
\cov\left[C_{\ell,\TE},C_{\ell,\TE}\right] & = \frac{1}{f_\mathrm{sky}}
\frac{1}{2\ell+1} \left[C_{\ell,\TE}^2+\right.\nn \\
 &\quad\left. (C_{\ell,\EE}+N_{\ell,\TT})(C_{\ell,\EE}+N_{\ell,\EE})\right]\ .
\end{align}
In the above $f_\mathrm{sky}$ is the fraction of sky that is surveyed and
usable for science, and we take $f_\mathrm{sky}=0.8$.  The terms
$N_{\ell,\TT}$ and $N_{\ell,\EE}$ denote the beam-noise in the temperature
and polarization detectors, respectively. These can be expressed as:
\begin{align} 
N_{\ell,\TT} & = \left[w_{\TT}W^2_\mathrm{Beam}(\ell)\right]^{-1}\\
N_{\ell,\EE} & =  \left[w_{\EE}W^2_\mathrm{Beam}(\ell)\right]^{-1} \ ,
\end{align}
where $w_{\TT}=\left[\Delta_\mathrm{T}\theta_\mathrm{Beam}\right]^{-1}$ and
$w_{\EE}=\left[\Delta_\mathrm{E}\theta_\mathrm{Beam}\right]^{-1}$. The beam
window function has the form:
\begin{equation}
W^2_\mathrm{Beam}(\ell)=\exp\left[-\ell(\ell+1)\sigma^2_\mathrm{Beam}\right]; \quad
\sigma_\mathrm{Beam}\equiv \frac{\theta_\mathrm{Beam}}{\sqrt{8\log 2}}\ .
\end{equation}
For the Planck experiment we assume that we have a single frequency
band for science (143 GHz), and for this channel the following
parameters apply \citep{PlanckBlueBook}: angular resolution of the
beam $\theta_\mathrm{Beam}=7.1'$ [FWHM]; the beam intensity is
\mbox{$\Delta_\mathrm{T}=2.2\, (T_\mathrm{CMB}/1\mathrm{K}) \, [\mu\rm K]$},
\mbox{$\Delta_\mathrm{E}=4.2\, (T_\mathrm{CMB}/1\mathrm{K}) \, [\mu\rm K]$}. We
take the temperature of the CMB to be $T=2.726 \mathrm{K}$.

\subsection{Transforming from CMB to Weak Lensing parameters}

In the lensing simulations, we have considered how the weak
lensing signal depends on the parameters,
$\{w_0,\Omega_\mathrm{m},n_\mathrm{s},\sigma_8,\fNL\}$. Let us rewrite our
original 10-D parameter set in terms of a new 10-D parameters
set. Therefore let us consider the transformation:
\begin{equation}
\begin{split}
\vect{p} & = \transposed{(w_0,w_a,\Omega_\mathrm{DE},w_\mathrm{c},w_\mathrm{b},\tau,n_\mathrm{s},A_\mathrm{s},\alpha,\fNL)} \\
\mapsto
\vect{q} & = \transposed{(w_0,w_a,\Omega_\mathrm{m},h,f_b,\tau,n_\mathrm{s},\sigma_8,\alpha,\fNL)} .
\end{split}
\end{equation}
Six of the parameters are unchanged from the original set.
The remaining four are related to the original parameters in the
following way:
\begin{align}
\Omega_\mathrm{m} & = 1-\Omega_\mathrm{DE} \ ;\label{eq:par1}\\
f_b & = \frac{w_\mathrm{b}}{w_\mathrm{b}+w_\mathrm{c}}\ ;\label{eq:par2}\\
h & = \sqrt{\frac{w_\mathrm{b}+w_\mathrm{c}}{1-\Omega_\mathrm{DE}}}\ ;\label{eq:par3} \\
\sigma_8 & = \left[\int \frac{\dk}{(2\pi)^3}P(k|\vect{p})
  W(kR)^2\right]^{1/2} , \label{eq:par4}
\end{align}
where $P(k|\vect{p})$ is the matter power spectrum, which depends on
parameters $p_{\alpha}$, and where the real space, spherical top-hat
filter function has the form $W_k(y)=3[\sin y-y \cos y]/y^3$, with
$y\equiv kR$ and $R=8\Mpc$. 

We now need to construct the matrix $\matrb{\Lambda}$, i.e. the partial
derivatives of the old parameters with respect to the new
parameters. From Eqns~(\ref{eq:par1})--(\ref{eq:par4}) we have
$q_a=G_a(\vect{p})$, however in order to perform the partial
derivatives we actually require the inverse of these relations, i.e.
$p_a=G_a^{-1}(\vect{q})$. In some cases these inverse relations may
easily be determined, e.g. \Eqn{eq:par1}. However, in other cases no
analytic inverse exists, e.g. \Eqn{eq:par4}.  A simple way around this
problem is through recalling the following:
\begin{equation} 
\sum_{\alpha}\frac{\partial p_{\mu}}{\partial q_{\alpha}}\frac{\partial q_{\alpha}}{\partial p_{\nu}}
=\delta^{K}_{\mu\nu}
\end{equation}
Or in other terms,
\begin{equation} \sum_{\alpha}\Lambda_{\mu\alpha}\frac{\partial
  q_{\alpha}}{\partial p_{\nu}}=
\sum_{\alpha}\Lambda_{\mu\alpha}\Lambda^{-1}_{\alpha\nu}=\delta^{K}_{\mu\nu}
\end{equation}
Hence, with ${\partial q_{\alpha}}/{\partial
  p_{\nu}}\equiv\Lambda^{-1}_{\alpha,\nu}$, we may then simply compute
$\matrb{\Lambda}$, through: $\matrb{\Lambda} = \bigl(\matrb{\Lambda}^{-1}\bigr)^{-1}$.

Let us therefore form the matrix $\Lambda^{-1}$. For the case of the
five untransformed parameters, we have
$\Lambda^{-1}_{\alpha\beta}=\delta^{K}_{\alpha\beta}$. However, for
the other four parameters, the situation is more complex and the
%
non-trivial elements are:
\begin{align}
\frac{\partial \Omega_\mathrm{m}}{\partial \Omega_\mathrm{DE}} & =  -1 \ ; \\
\frac{\partial f_b}{\partial w_\mathrm{c}} & =  -\frac{w_\mathrm{b}}{[w_\mathrm{c}+w_\mathrm{b}]^2} \ ; \\ 
\frac{\partial f_b}{\partial w_\mathrm{b}} & =  \frac{w_\mathrm{c}}{[w_\mathrm{c}+w_\mathrm{b}]^2} \ ; \\ 
\frac{\partial h}{\partial \Omega_\mathrm{DE}} & = \frac{1}{2}\sqrt{\frac{w_\mathrm{c}+w_\mathrm{b}}{(1-\Omega_\mathrm{DE})^3}} \ ; \\ 
\frac{\partial h}{\partial w_\mathrm{c}} & =  \frac{1}{2}\left[(1-\Omega_\mathrm{DE})(w_\mathrm{b}+w+c)\right]^{-1/2} \ ; \\
\frac{\partial \sigma_8}{\partial p_{\alpha}} & = 
\frac{1}{2\sigma_8} \frac{\partial \sigma_8^2}{\partial p_{\alpha}} = 
\frac{1}{2\sigma_8} \int \frac{\dk}{(2\pi)^3} \frac{\partial P(k|\vect{p})}{\partial p_{\alpha}} W(kR)^2\ .
\end{align}

Note, at leading order in $\fNL$ we have, $\partial \sigma_8/\partial \fNL=0$. In order to compute the derivatives ${\partial \sigma_8}/{\partial p_{\alpha}}$ we use the package CAMB. Besides the generation of various CMB power spectra, this package can output the present day linear theory matter power spectra $P(k|\vect{p})$. The derivatives are then determined numerically using the standard estimator for two sided derivatives.

\begin{table}
\caption{
\label{tab:cov_matrix_CMB}
Covariance $\matrb{C}_\text{pr}$ of the non-Gaussianity parameter $\fNL$, the mean matter density $\Omega_\text{m}$, the normalization $\sigma_8$, the spectral index $n_\text{s}$, and the equation-of-state parameter $w_0$ assumed for a prior parameter distribution provided by a CMB experiment like \satellitename{Planck}.
}
\begin{tabular}{l r r r r r}
                  & $\fNL$ & $\Omega_\text{m}$ & $\sigma_8$ & $n_\text{s}$ & $w_0$ \\
\hline 
$\!\!\fNL           \!\!$ & $\!\! \infty                 \!\!$ & $\!\! 0                      \!\!$ & $\!\! 0                      \!\!$ & $\!\! 0                      \!\!$ & $\!\! 0                      \!\!$ \\
$\!\!\Omega_\text{m}\!\!$ & $\!\! 0                      \!\!$ & $\!\! 2.2 \!\times\! 10^{ -2}\!\!$ & $\!\!-3.5 \!\times\! 10^{ -2}\!\!$ & $\!\!-1.2 \!\times\! 10^{ -5}\!\!$ & $\!\! 1.0 \!\times\! 10^{ -1}\!\!$ \\
$\!\!\sigma_8       \!\!$ & $\!\! 0                      \!\!$ & $\!\!-3.5 \!\times\! 10^{ -2}\!\!$ & $\!\! 5.9 \!\times\! 10^{ -2}\!\!$ & $\!\!-8.9 \!\times\! 10^{ -6}\!\!$ & $\!\!-1.7 \!\times\! 10^{ -1}\!\!$ \\
$\!\!n_\text{s}     \!\!$ & $\!\! 0                      \!\!$ & $\!\!-1.2 \!\times\! 10^{ -5}\!\!$ & $\!\!-8.9 \!\times\! 10^{ -6}\!\!$ & $\!\! 1.9 \!\times\! 10^{ -5}\!\!$ & $\!\! 2.8 \!\times\! 10^{ -6}\!\!$ \\
$\!\!w_0            \!\!$ & $\!\! 0                      \!\!$ & $\!\! 1.0 \!\times\! 10^{ -1}\!\!$ & $\!\!-1.7 \!\times\! 10^{ -1}\!\!$ & $\!\! 2.8 \!\times\! 10^{ -6}\!\!$ & $\!\! 5.8 \!\times\! 10^{ -1}\!\!$ \\
           
\end{tabular}       
\end{table}         

The numerical inverse of the matrix $\matrb{\Lambda}^{-1}$ can easily be computed using the SVD algorithm \citep{PressEtal1992}. We can then transform the CMB Fisher matrix from the original to the new parameter set. The final step is a marginalization over those parameters not considered in our weak-lensing analysis.
The covariances for the remaining parameters $\fNL$, $\Omega_\mathrm{m}$, $\sigma_8$, $n_\mathrm{s}$, and $w_0$ are shown in Table~\ref{tab:cov_matrix_CMB}.

\section{Parameter covariance matrices}
\label{sec:appendix:parameter_covariance_matrices}

\begin{table}
\caption{
\label{tab:cov_matrix_xi_pm}
Posterior covariance of the non-Gaussianity parameter $\fNL$, the mean matter density $\Omega_\text{m}$, the normalization $\sigma_8$, the spectral index $n_\text{s}$, and the equation-of-state parameter $w_0$ expected from measurements of the cosmic shear correlation functions $\xi_\pm$. 
}
\begin{tabular}{l r r r r r}
                  & $\fNL$ & $\Omega_\text{m}$ & $\sigma_8$ & $n_\text{s}$ & $w_0$ \\
\hline 
$\!\!\fNL           \!\!$ & $\!\! 4.5 \!\times\! 10^{  4}\!\!$ & $\!\!-5.1 \!\times\! 10^{ -1}\!\!$ & $\!\!-2.5 \!\times\! 10^{  0}\!\!$ & $\!\!-1.0 \!\times\! 10^{ -1}\!\!$ & $\!\!-1.0 \!\times\! 10^{  1}\!\!$ \\
$\!\!\Omega_\text{m}\!\!$ & $\!\!-5.1 \!\times\! 10^{ -1}\!\!$ & $\!\! 4.0 \!\times\! 10^{ -5}\!\!$ & $\!\! 3.0 \!\times\! 10^{ -5}\!\!$ & $\!\!-1.4 \!\times\! 10^{ -4}\!\!$ & $\!\! 6.9 \!\times\! 10^{ -4}\!\!$ \\
$\!\!\sigma_8       \!\!$ & $\!\!-2.5 \!\times\! 10^{  0}\!\!$ & $\!\! 3.0 \!\times\! 10^{ -5}\!\!$ & $\!\! 1.4 \!\times\! 10^{ -4}\!\!$ & $\!\!-1.4 \!\times\! 10^{ -5}\!\!$ & $\!\! 6.2 \!\times\! 10^{ -4}\!\!$ \\
$\!\!n_\text{s}     \!\!$ & $\!\!-1.0 \!\times\! 10^{ -1}\!\!$ & $\!\!-1.4 \!\times\! 10^{ -4}\!\!$ & $\!\!-1.4 \!\times\! 10^{ -5}\!\!$ & $\!\! 6.7 \!\times\! 10^{ -4}\!\!$ & $\!\!-2.4 \!\times\! 10^{ -3}\!\!$ \\
$\!\!w_0            \!\!$ & $\!\!-1.0 \!\times\! 10^{  1}\!\!$ & $\!\! 6.9 \!\times\! 10^{ -4}\!\!$ & $\!\! 6.2 \!\times\! 10^{ -4}\!\!$ & $\!\!-2.4 \!\times\! 10^{ -3}\!\!$ & $\!\! 1.3 \!\times\! 10^{ -2}\!\!$ \\
           
\end{tabular}       
\end{table}         

\begin{table}
\center
\caption{
\label{tab:cov_matrix_map}
Posterior parameter covariances expected from measurements of the third-order aperture statistics $\bEV{M_\text{ap}^3}$. 
}
\begin{tabular}{l r r r r r}
                  & $\fNL$ & $\Omega_\text{m}$ & $\sigma_8$ & $n_\text{s}$ & $w_0$ \\
\hline 
$\!\!\fNL           \!\!$ &  $\!\! 2.0 \!\times\! 10^{  5}\!\!$ & $\!\! 6.5 \!\times\! 10^{ -1}\!\!$ & $\!\!-1.8 \!\times\! 10^{  1}\!\!$ & $\!\!-1.1 \!\times\! 10^{  1}\!\!$ & $\!\! 1.7 \!\times\! 10^{  1}\!\!$ \\
$\!\!\Omega_\text{m}\!\!$ &  $\!\! 6.5 \!\times\! 10^{ -1}\!\!$ & $\!\! 1.0 \!\times\! 10^{ -4}\!\!$ & $\!\!-1.7 \!\times\! 10^{ -4}\!\!$ & $\!\!-4.0 \!\times\! 10^{ -4}\!\!$ & $\!\! 1.6 \!\times\! 10^{ -3}\!\!$ \\
$\!\!\sigma_8       \!\!$ &  $\!\!-1.8 \!\times\! 10^{  1}\!\!$ & $\!\!-1.7 \!\times\! 10^{ -4}\!\!$ & $\!\! 1.8 \!\times\! 10^{ -3}\!\!$ & $\!\! 1.4 \!\times\! 10^{ -3}\!\!$ & $\!\!-3.2 \!\times\! 10^{ -3}\!\!$ \\
$\!\!n_\text{s}     \!\!$ &  $\!\!-1.1 \!\times\! 10^{  1}\!\!$ & $\!\!-4.0 \!\times\! 10^{ -4}\!\!$ & $\!\! 1.4 \!\times\! 10^{ -3}\!\!$ & $\!\! 2.1 \!\times\! 10^{ -3}\!\!$ & $\!\!-6.9 \!\times\! 10^{ -3}\!\!$ \\
$\!\!w_0            \!\!$ &  $\!\! 1.7 \!\times\! 10^{  1}\!\!$ & $\!\! 1.6 \!\times\! 10^{ -3}\!\!$ & $\!\!-3.2 \!\times\! 10^{ -3}\!\!$ & $\!\!-6.9 \!\times\! 10^{ -3}\!\!$ & $\!\! 3.5 \!\times\! 10^{ -2}\!\!$ \\
           
\end{tabular}       
\end{table}         

\begin{table}
\center
\caption{
\label{tab:cov_matrix_dpa}
Posterior parameter covariances expected from measurements of the shear peak abundance $\dpa$.
}
\begin{tabular}{l r r r r r}
                  & $\fNL$ & $\Omega_\text{m}$ & $\sigma_8$ & $n_\text{s}$ & $w_0$ \\
\hline 
$\!\!\fNL           \!\!$ & $\!\! 4.8 \!\times\! 10^{  3}\!\!$ & $\!\! 1.0 \!\times\! 10^{ -1}\!\!$ & $\!\!-5.6 \!\times\! 10^{ -1}\!\!$ & $\!\! 2.3 \!\times\! 10^{ -1}\!\!$ & $\!\! 3.5 \!\times\! 10^{ -1}\!\!$ \\
$\!\!\Omega_\text{m}\!\!$ & $\!\! 1.0 \!\times\! 10^{ -1}\!\!$ & $\!\! 7.0 \!\times\! 10^{ -5}\!\!$ & $\!\!-1.1 \!\times\! 10^{ -4}\!\!$ & $\!\! 4.3 \!\times\! 10^{ -5}\!\!$ & $\!\! 8.2 \!\times\! 10^{ -4}\!\!$ \\
$\!\!\sigma_8       \!\!$ & $\!\!-5.6 \!\times\! 10^{ -1}\!\!$ & $\!\!-1.1 \!\times\! 10^{ -4}\!\!$ & $\!\! 2.2 \!\times\! 10^{ -4}\!\!$ & $\!\!-1.6 \!\times\! 10^{ -4}\!\!$ & $\!\!-1.4 \!\times\! 10^{ -3}\!\!$ \\
$\!\!n_\text{s}     \!\!$ & $\!\! 2.3 \!\times\! 10^{ -1}\!\!$ & $\!\! 4.3 \!\times\! 10^{ -5}\!\!$ & $\!\!-1.6 \!\times\! 10^{ -4}\!\!$ & $\!\! 1.0 \!\times\! 10^{ -3}\!\!$ & $\!\! 2.4 \!\times\! 10^{ -3}\!\!$ \\
$\!\!w_0            \!\!$ & $\!\! 3.5 \!\times\! 10^{ -1}\!\!$ & $\!\! 8.2 \!\times\! 10^{ -4}\!\!$ & $\!\!-1.4 \!\times\! 10^{ -3}\!\!$ & $\!\! 2.4 \!\times\! 10^{ -3}\!\!$ & $\!\! 1.4 \!\times\! 10^{ -2}\!\!$ \\
           
\end{tabular}       
\end{table}         

\begin{table}
\center
\caption{
\label{tab:cov_matrix_xp_pm_and_map}
Posterior parameter covariances expected from joint measurements of the cosmic shear correlation functions $\xi_\pm$ and the aperture statistics $\bEV{M_\text{ap}^3}$. 
}
\begin{tabular}{l r r r r r}
                  & $\fNL$ & $\Omega_\text{m}$ & $\sigma_8$ & $n_\text{s}$ & $w_0$ \\
\hline 
$\!\!\fNL           \!\!$ & $\!\! 2.0 \!\times\! 10^{  4}\!\!$ & $\!\!-5.6 \!\times\! 10^{ -1}\!\!$ & $\!\!-1.1 \!\times\! 10^{  0}\!\!$ & $\!\! 1.4 \!\times\! 10^{  0}\!\!$ & $\!\!-1.1 \!\times\! 10^{  1}\!\!$ \\
$\!\!\Omega_\text{m}\!\!$ & $\!\!-5.6 \!\times\! 10^{ -1}\!\!$ & $\!\! 1.8 \!\times\! 10^{ -5}\!\!$ & $\!\! 3.1 \!\times\! 10^{ -5}\!\!$ & $\!\!-4.8 \!\times\! 10^{ -5}\!\!$ & $\!\! 3.2 \!\times\! 10^{ -4}\!\!$ \\
$\!\!\sigma_8       \!\!$ & $\!\!-1.1 \!\times\! 10^{  0}\!\!$ & $\!\! 3.1 \!\times\! 10^{ -5}\!\!$ & $\!\! 6.8 \!\times\! 10^{ -5}\!\!$ & $\!\!-8.5 \!\times\! 10^{ -5}\!\!$ & $\!\! 6.0 \!\times\! 10^{ -4}\!\!$ \\
$\!\!n_\text{s}     \!\!$ & $\!\! 1.4 \!\times\! 10^{  0}\!\!$ & $\!\!-4.8 \!\times\! 10^{ -5}\!\!$ & $\!\!-8.5 \!\times\! 10^{ -5}\!\!$ & $\!\! 1.6 \!\times\! 10^{ -4}\!\!$ & $\!\!-8.2 \!\times\! 10^{ -4}\!\!$ \\
$\!\!w_0            \!\!$ & $\!\!-1.1 \!\times\! 10^{  1}\!\!$ & $\!\! 3.2 \!\times\! 10^{ -4}\!\!$ & $\!\! 6.0 \!\times\! 10^{ -4}\!\!$ & $\!\!-8.2 \!\times\! 10^{ -4}\!\!$ & $\!\! 5.9 \!\times\! 10^{ -3}\!\!$ \\
           
\end{tabular}       
\end{table}         
  
\begin{table}
\center
\caption{
\label{tab:cov_matrix_xi_pm_and_dpa}
Posterior parameter covariances expected from joint measurements of the cosmic shear correlation functions $\xi_\pm$ and the shear peak abundance $\dpa$.
}
\begin{tabular}{l r r r r r}
                  & $\fNL$ & $\Omega_\text{m}$ & $\sigma_8$ & $n_\text{s}$ & $w_0$ \\
\hline 
$\!\!\fNL           \!\!$ & $\!\! 3.6 \!\times\! 10^{  3}\!\!$ & $\!\!-8.2 \!\times\! 10^{ -2}\!\!$ & $\!\!-2.0 \!\times\! 10^{ -1}\!\!$ & $\!\! 2.2 \!\times\! 10^{ -1}\!\!$ & $\!\!-1.4 \!\times\! 10^{  0}\!\!$ \\
$\!\!\Omega_\text{m}\!\!$ & $\!\!-8.2 \!\times\! 10^{ -2}\!\!$ & $\!\! 3.9 \!\times\! 10^{ -6}\!\!$ & $\!\! 2.5 \!\times\! 10^{ -6}\!\!$ & $\!\!-1.1 \!\times\! 10^{ -5}\!\!$ & $\!\! 4.5 \!\times\! 10^{ -5}\!\!$ \\
$\!\!\sigma_8       \!\!$ & $\!\!-2.0 \!\times\! 10^{ -1}\!\!$ & $\!\! 2.5 \!\times\! 10^{ -6}\!\!$ & $\!\! 1.5 \!\times\! 10^{ -5}\!\!$ & $\!\!-1.2 \!\times\! 10^{ -5}\!\!$ & $\!\! 6.3 \!\times\! 10^{ -5}\!\!$ \\
$\!\!n_\text{s}     \!\!$ & $\!\! 2.2 \!\times\! 10^{ -1}\!\!$ & $\!\!-1.1 \!\times\! 10^{ -5}\!\!$ & $\!\!-1.2 \!\times\! 10^{ -5}\!\!$ & $\!\! 7.1 \!\times\! 10^{ -5}\!\!$ & $\!\!-8.6 \!\times\! 10^{ -5}\!\!$ \\
$\!\!w_0            \!\!$ & $\!\!-1.4 \!\times\! 10^{  0}\!\!$ & $\!\! 4.5 \!\times\! 10^{ -5}\!\!$ & $\!\! 6.3 \!\times\! 10^{ -5}\!\!$ & $\!\!-8.6 \!\times\! 10^{ -5}\!\!$ & $\!\! 7.7 \!\times\! 10^{ -4}\!\!$ \\
           
\end{tabular}       
\end{table}         

\begin{table}
\center
\caption{
\label{tab:cov_matrix_map_and_dpa}
Posterior parameter covariances expected from joint measurements of the aperture statistics $\bEV{M_\text{ap}^3}$ and the shear peak abundance $\dpa$.
}
\begin{tabular}{l r r r r r}
                  & $\fNL$ & $\Omega_\text{m}$ & $\sigma_8$ & $n_\text{s}$ & $w_0$ \\
\hline 
$\!\!\fNL           \!\!$ & $\!\! 4.0 \!\times\! 10^{  3}\!\!$ & $\!\! 1.8 \!\times\! 10^{ -2}\!\!$ & $\!\!-3.3 \!\times\! 10^{ -1}\!\!$ & $\!\!-2.6 \!\times\! 10^{ -1}\!\!$ & $\!\!-1.4 \!\times\! 10^{  0}\!\!$ \\
$\!\!\Omega_\text{m}\!\!$ & $\!\! 1.8 \!\times\! 10^{ -2}\!\!$ & $\!\! 1.6 \!\times\! 10^{ -5}\!\!$ & $\!\!-2.0 \!\times\! 10^{ -5}\!\!$ & $\!\!-4.5 \!\times\! 10^{ -5}\!\!$ & $\!\! 9.0 \!\times\! 10^{ -5}\!\!$ \\
$\!\!\sigma_8       \!\!$ & $\!\!-3.3 \!\times\! 10^{ -1}\!\!$ & $\!\!-2.0 \!\times\! 10^{ -5}\!\!$ & $\!\! 5.3 \!\times\! 10^{ -5}\!\!$ & $\!\! 6.1 \!\times\! 10^{ -5}\!\!$ & $\!\!-4.2 \!\times\! 10^{ -6}\!\!$ \\
$\!\!n_\text{s}     \!\!$ & $\!\!-2.6 \!\times\! 10^{ -1}\!\!$ & $\!\!-4.5 \!\times\! 10^{ -5}\!\!$ & $\!\! 6.1 \!\times\! 10^{ -5}\!\!$ & $\!\! 2.5 \!\times\! 10^{ -4}\!\!$ & $\!\!-1.2 \!\times\! 10^{ -5}\!\!$ \\
$\!\!w_0            \!\!$ & $\!\!-1.4 \!\times\! 10^{  0}\!\!$ & $\!\! 9.0 \!\times\! 10^{ -5}\!\!$ & $\!\!-4.2 \!\times\! 10^{ -6}\!\!$ & $\!\!-1.2 \!\times\! 10^{ -5}\!\!$ & $\!\! 1.5 \!\times\! 10^{ -3}\!\!$ \\
           
\end{tabular}       
\end{table}         

\begin{table}
\center
\caption{
\label{tab:cov_matrix_full}
Posterior parameter covariances expected from joint measurements of the cosmic shear correlation functions $\xi_\pm$, the aperture statistics $\bEV{M_\text{ap}^3}$, and the shear peak abundance $\dpa$.
}
\begin{tabular}{l r r r r r}
                  & $\fNL$ & $\Omega_\text{m}$ & $\sigma_8$ & $n_\text{s}$ & $w_0$ \\
\hline 
$\!\!\fNL           \!\!$ & $\!\! 2.8 \!\times\! 10^{  3}\!\!$ & $\!\!-5.7 \!\times\! 10^{ -2}\!\!$ & $\!\!-1.6 \!\times\! 10^{ -1}\!\!$ & $\!\! 1.1 \!\times\! 10^{ -1}\!\!$ & $\!\!-1.2 \!\times\! 10^{  0}\!\!$ \\
$\!\!\Omega_\text{m}\!\!$ & $\!\!-5.7 \!\times\! 10^{ -2}\!\!$ & $\!\! 2.8 \!\times\! 10^{ -6}\!\!$ & $\!\! 1.7 \!\times\! 10^{ -6}\!\!$ & $\!\!-6.7 \!\times\! 10^{ -6}\!\!$ & $\!\! 3.5 \!\times\! 10^{ -5}\!\!$ \\
$\!\!\sigma_8       \!\!$ & $\!\!-1.6 \!\times\! 10^{ -1}\!\!$ & $\!\! 1.7 \!\times\! 10^{ -6}\!\!$ & $\!\! 1.3 \!\times\! 10^{ -5}\!\!$ & $\!\!-7.3 \!\times\! 10^{ -6}\!\!$ & $\!\! 5.9 \!\times\! 10^{ -5}\!\!$ \\
$\!\!n_\text{s}     \!\!$ & $\!\! 1.1 \!\times\! 10^{ -1}\!\!$ & $\!\!-6.7 \!\times\! 10^{ -6}\!\!$ & $\!\!-7.3 \!\times\! 10^{ -6}\!\!$ & $\!\! 4.5 \!\times\! 10^{ -5}\!\!$ & $\!\!-5.7 \!\times\! 10^{ -5}\!\!$ \\
$\!\!w_0            \!\!$ & $\!\!-1.2 \!\times\! 10^{  0}\!\!$ & $\!\! 3.5 \!\times\! 10^{ -5}\!\!$ & $\!\! 5.9 \!\times\! 10^{ -5}\!\!$ & $\!\!-5.7 \!\times\! 10^{ -5}\!\!$ & $\!\! 6.9 \!\times\! 10^{ -4}\!\!$ \\
           
\end{tabular}       
\end{table}         

\begin{table}
\center
\caption{
\label{tab:cov_matrix_full_and_CMB}
Posterior parameter covariances expected from joint measurements of the cosmic shear correlation functions $\xi_\pm$, the aperture statistics $\bEV{M_\text{ap}^3}$, and the shear peak abundance $\dpa$, when combined with prior information from a CMB experiment like \satellitename{Planck}.
}
\begin{tabular}{l r r r r r}
                  & $\fNL$ & $\Omega_\text{m}$ & $\sigma_8$ & $n_\text{s}$ & $w_0$ \\
\hline 
$\!\!\fNL           \!\!$ & $\!\! 2.6 \!\times\! 10^{  3}\!\!$ & $\!\!-4.5 \!\times\! 10^{ -2}\!\!$ & $\!\!-1.4 \!\times\! 10^{ -1}\!\!$ & $\!\! 3.3 \!\times\! 10^{ -2}\!\!$ & $\!\!-1.1 \!\times\! 10^{  0}\!\!$ \\
$\!\!\Omega_\text{m}\!\!$ & $\!\!-4.5 \!\times\! 10^{ -2}\!\!$ & $\!\! 2.1 \!\times\! 10^{ -6}\!\!$ & $\!\! 9.0 \!\times\! 10^{ -7}\!\!$ & $\!\!-2.0 \!\times\! 10^{ -6}\!\!$ & $\!\! 2.9 \!\times\! 10^{ -5}\!\!$ \\
$\!\!\sigma_8       \!\!$ & $\!\!-1.4 \!\times\! 10^{ -1}\!\!$ & $\!\! 9.0 \!\times\! 10^{ -7}\!\!$ & $\!\! 1.2 \!\times\! 10^{ -5}\!\!$ & $\!\!-2.2 \!\times\! 10^{ -6}\!\!$ & $\!\! 5.1 \!\times\! 10^{ -5}\!\!$ \\
$\!\!n_\text{s}     \!\!$ & $\!\! 3.3 \!\times\! 10^{ -2}\!\!$ & $\!\!-2.0 \!\times\! 10^{ -6}\!\!$ & $\!\!-2.2 \!\times\! 10^{ -6}\!\!$ & $\!\! 1.3 \!\times\! 10^{ -5}\!\!$ & $\!\!-1.7 \!\times\! 10^{ -5}\!\!$ \\
$\!\!w_0            \!\!$ & $\!\!-1.1 \!\times\! 10^{  0}\!\!$ & $\!\! 2.9 \!\times\! 10^{ -5}\!\!$ & $\!\! 5.1 \!\times\! 10^{ -5}\!\!$ & $\!\!-1.7 \!\times\! 10^{ -5}\!\!$ & $\!\! 6.4 \!\times\! 10^{ -4}\!\!$ \\
           
\end{tabular}       
\end{table}         

Tables \ref{tab:cov_matrix_xi_pm}-\ref{tab:cov_matrix_full_and_CMB} show the posterior parameter covariances expected from the measurements of the cosmic shear correlation functions $\xi_\pm$, the aperture statistics $\bEV{M_\text{ap}^3}$ and the shear peak abundance $\dpa$ in a large future weak lensing survey. Electronic versions of these tables are available online.

\end{document}